\documentclass{emulateapj}

\newcommand{\ldl}{$\lambda/{\Delta}{\lambda}$}
\newcommand{\teff}{T$_{eff}$}
\newcommand{\kion}{\ion{K}{1}}
\newcommand{\naion}{\ion{Na}{1}}
\newcommand{\meth}{CH$_4$}
\newcommand{\water}{H$_2$O}

\slugcomment{Accepted for publication in ApJ}

\shorttitle{The Physical Properties of T Dwarfs}
\shortauthors{Burgasser, Burrows \& Kirkpatrick}

\begin{document}

\title{A Method for Determining the Physical Properties of the Coldest Known Brown Dwarfs}

\author{Adam J.\ Burgasser\altaffilmark{1}}
\affil{Massachusetts Institute of Technology, Kavli Institute for Astrophysics and Space Research,
77 Massachusetts Avenue, Building 37,
Cambridge, MA 02139-4307, USA; ajb@mit.edu}
\author{Adam Burrows}
\affil{Department of Astronomy, University of Arizona,
Tucson, AZ 85721, USA; burrows@as.arizona.edu}
\and
\author{J.\ Davy Kirkpatrick}
\affil{Infrared Processing and Analysis Center, M/S 100-22,
California Institute of Technology, Pasadena, CA 91125, USA; davy@ipac.caltech.edu}

\altaffiltext{1}{Visiting Astronomer at the Infrared Telescope Facility, which is operated by
the University of Hawaii under Cooperative Agreement NCC 5-538 with the National Aeronautics
and Space Administration, Office of Space Science, Planetary Astronomy Program.}

\begin{abstract}
We present a method for measuring the physical parameters
of the coldest T-type brown dwarfs using low resolution near infrared spectra.
By comparing {\water}- and H$_2$-sensitive spectral ratios
between empirical data and theoretical atmosphere models,
and calibrating these ratios
to measurements for the well-characterized 2--5 Gyr companion brown dwarf Gliese~570D,
we derive estimates of the effective temperatures and surface gravities
for 13 mid- and late-type field T dwarfs.  We also deduce the first quantitative estimate
of subsolar metallicity for the peculiar T dwarf 2MASS~0937+2931.
Derived temperatures
are consistent with prior estimates based on parallax and
bolometric luminosity measurements, and examination of possible
systematic effects indicate that the results are robust.
Two recently discovered late-type T dwarfs,
2MASS~0939$-$2448 and 2MASS~1114$-$2618, both appear to be
$\gtrsim$50~K cooler than the latest-type T dwarf, 2MASS~0415-0935,
and are potentially the coldest and least luminous
brown dwarfs currently known.
We find that, in general, higher surface gravity T dwarfs
have lower effective temperatures and luminosities
for a given spectral type,
explaining previously observed scatter in
the {\teff}/spectral type relation for these objects.
Masses, radii and ages are estimated for the T dwarfs in our sample
using the evolutionary models of Burrows et al.; we
also determine masses and radii independently
for eight T dwarfs with measured luminosities.
These two determinations are largely
consistent, lending support to the validity of evolutionary models at late ages.
Our method is well suited to large samples of faint
brown dwarfs, and can ultimately be used to directly measure the
substellar mass function and formation history in the Galaxy.
\end{abstract}


\section{Introduction}

The spectral energy distributions of the coldest known
stars and brown dwarfs, L dwarfs and T dwarfs \citep{kir99,me02a,geb02},
are complex, dominated by
broad, overlapping gaseous and condensate molecular absorption features.
The strengths of these features depend on a combination of photospheric
temperature, gas pressure and composition (e.g., Burrows \& Sharp 1999; Lodders \& Fegley 2002),
%\citep{bur99,lod02},
which in turn are related to the effective temperature ({\teff}), surface gravity ($g$)
and metallicity ([M/H]) of a brown dwarf.
Nonequilibrium effects (e.g., vertical mixing, cloud coverage) may also play an important
role in molecular \citep{feg96,sau03} and condensate \citep{ack01,me02c}
abundances.  The combined influence of these
parameters on the spectra of L and T dwarfs is only beginning to be
explored through the study of low mass, substellar objects in young clusters
and stellar associations
\citep{luc01,gor03,mcg04} and ultracool subdwarfs \citep{me03f,sch04}, although systematic
studies have yet to be achieved.

Disentangling the properties of {\teff}, surface gravity
and metallicity is a principal goal
of substellar astrophysics.  These parameters can be used to infer
masses, radii and ages for individual sources
(e.g., Mohanty, Jaywardhana \& Basri 2004), allowing, in the long term,
direct measurement of the substellar mass function (MF) and star formation
history for field objects in the Solar Neighborhood
\citep{cha03,me04a,all05}. In the short term, {\teff}, $g$ and [M/H] measurements
for young cluster or companion
brown dwarfs enable tests of evolutionary models \citep{moh04a}.

Gravity and metallicity effects are particularly relevant for interpreting
the spectral energy distributions of the coldest T dwarfs, spectral types T6 and later.
These objects, with {\teff} $\lesssim$ 1000 K \citep{gol04}, lack the complicating
influence of photospheric condensates common in late-type M dwarfs, L dwarfs
and the earliest-type T dwarfs \citep{tsu96,ack01,all01}.  They exhibit
good correlation between spectral type and {\teff}
\citep{dah02,tin03,gol04,nak04,vrb04}.
Surface gravity and metallicity effects are therefore readily
distinguished by the presence of spectral or photometric anomalies.
One case in point is the peculiar
T6 dwarf 2MASS~0937+2931\footnote{Source designations in this article are
abbreviated in the manner 2MASS~hhmm$\pm$ddmm;
the suffix is the sexagesimal Right
Ascension (hours and minutes) and declination (degrees and arcminutes)
at J2000 equinox. Full designations are provided
in Table~\ref{tab:sample}.} \citep{me02a}, a brown dwarf believed to have a high
surface gravity and/or subsolar metallicity \citep{me02a,me03d,bur02,kna04}.
2MASS~0937+2931 is 0.5--1.0 mag bluer than
similarly-classified T dwarfs; and its
spectrum exhibits a
suppressed $K$-band peak,
an extremely red 0.8--1.0 $\micron$ spectral slope, enhanced FeH absorption
at 0.99 $\micron$, and an absence of \ion{K}{1} doublet lines at 1.17 and 1.25 $\micron$,
all unusual for a mid-type T dwarf.
Several other late-type T dwarfs
exhibit similar color and spectral peculiarities
\citep{me03e,me04b,kna04}.
However, quantitative analysis of these effects, in the form of specific
surface gravity and metallicity measurements,
has been limited \citep{bur02,kna04}.

In this article, we present a method for disentangling {\teff}, $g$
and [M/H] effects in the near
infrared spectra of the latest-type T dwarfs.
Our method, based on the comparison of calibrated near infrared flux ratios
measured on low resolution spectral data
and theoretical models, yields strong constraints on
these physical parameters
and a means of estimating masses, radii and ages for individual field brown dwarfs.
In $\S$~2 we describe the sample and spectroscopic observations obtained
with the SpeX spectrograph \citep{ray03} mounted on the 3m NASA Infrared Telescope Facility (IRTF).
We identify and compare spectral variations observed in these low resolution near infrared spectra,
and discuss qualitatively how these features are associated
with differences in {\teff}, $g$
and [M/H].
In $\S$~3, we examine these same effects with theoretical models,
and characterize spectral trends.
In $\S$~4, we describe our method, and
present {\teff} and $\log{g}$ estimates for 13 field brown dwarfs and constraints
for two others; we also deduce subsolar metallicity estimates for two sources
including 2MASS~0937+2931.  We demonstrate the
consistency of our {\teff} values with previous determinations based on
parallax and luminosity measurements, and examine potential systematic effects.
In $\S$~5, we derive mass, radius and age estimates for our T dwarfs
using the evolutionary models of \citet{bur97}; and independently
determine masses and radii for eight sources with published
luminosity measurements.
We discuss the results in $\S$~6, focusing on new insights on the
{\teff}/spectral type relation for T dwarfs and potential
applications of our method for various brown dwarf studies.
Results are summarized in $\S$~7.

\section{Observations}

\subsection{The Sample}

Our primary spectral sample was composed of 16 T dwarfs
identified by \citet{str99}; \citet{me99,me02a,me04b}; \citet{tsv00};
\citet{geb02}; \citet{me03e}; \citet{kna04}; and \citet{tin05}
in the Two Micron All Sky Survey \citep[hereafter 2MASS]{cut03}
and the Sloan Digital Sky Survey \citep[hereafter SDSS]{yor00}.  The empirical
properties of these sources are listed in Table~\ref{tab:sample}.
The sample was selected to span types T5.5 to T8, based
on the unified classification scheme of \citet{me05c}, and excludes known
binaries \citep[Burgasser et al.\ in prep.]{me03b}.
Eight of these objects have parallax measurements from \citet{dah02,tin03};
and \citet{vrb04}; all but one has a reported proper motion.
Apparent 2MASS $J$-band magnitudes for these sources range from
14.7 to 16.3 mag.

\subsection{Near Infrared Spectroscopy}

Six of the T dwarfs in our sample -- 2MASS~0034+0523, 2MASS~0243-2453,
2MASS~0415-0935, 2MASS~1231+0847, Gliese~570D and 2MASS~2228-4310 --
have been previously observed with SpeX \citep{me04b}.
The remaining sources were observed during three runs on
2004 March 11--12, 2004 July 23 and 2004 September 7 (UT).
A log of observations is provided in Table~\ref{tab:spexlog}.
Conditions during the March run were
clear and dry with typical seeing of 0$\farcs$7.
Conditions during July were also clear with excellent seeing
(0$\farcs$4--0$\farcs$7).  Light cirrus was present during
the September observations, but seeing was again excellent
(0$\farcs$5--0$\farcs$7).

Spectral data for all of the sources in our sample
(including those previously observed)
were obtained using the SpeX prism dispersed mode,
which provides low resolution 0.7--2.5 $\micron$ spectra in a single order.
This setting minimizes spectral color errors commonly
incurred through order stitching (e.g., McLean et al.\ 2003),
yielding an accurate measure of the broad band spectral energy distribution.
For all observations, the 0$\farcs$5 slit was employed and rotated
to the parallactic angle, resulting in a
spectral resolution {\ldl} $\approx 150$
and dispersion across the chip of 20--30 {\AA} pixel$^{-1}$.
Multiple exposures of 180 s were obtained in an ABBA dither pattern along the slit.
Flux calibration
was made through observations of nearby A0 V stars obtained immediately
before or after the target observation and at similar airmasses ($\Delta\sec{z} < 0.1$).
Internal flat field and Ar arc lamps were observed after each flux calibrator star
for pixel response and wavelength calibration.
All spectral data were reduced using the SpeXtool package version 3.2
\citep{vac03,cus04} using standard settings.
Further details on the experimental design and
data reduction are given in \citet{me04b}.

\begin{figure}
\epsscale{1.0}
\plottwo{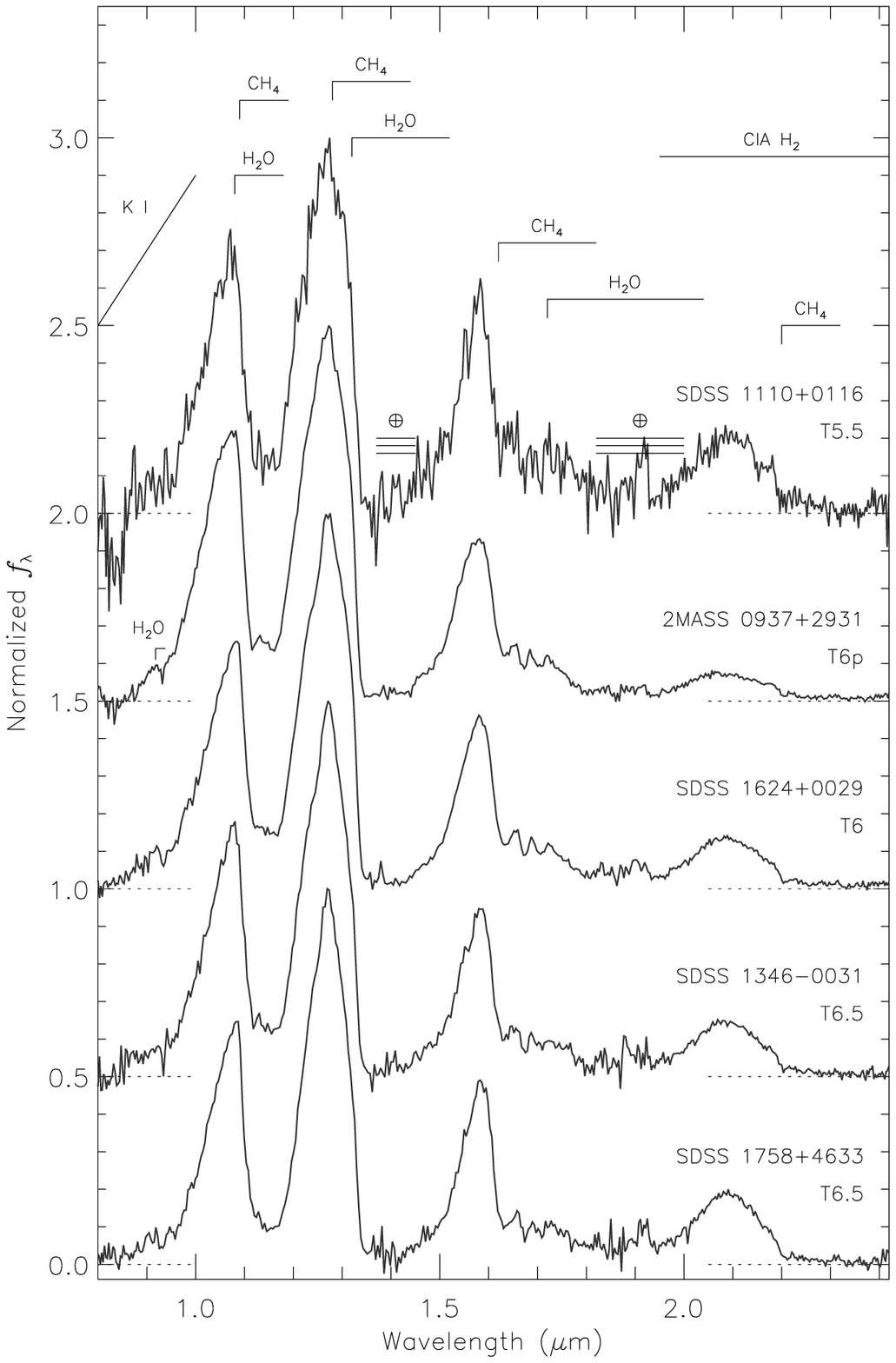}{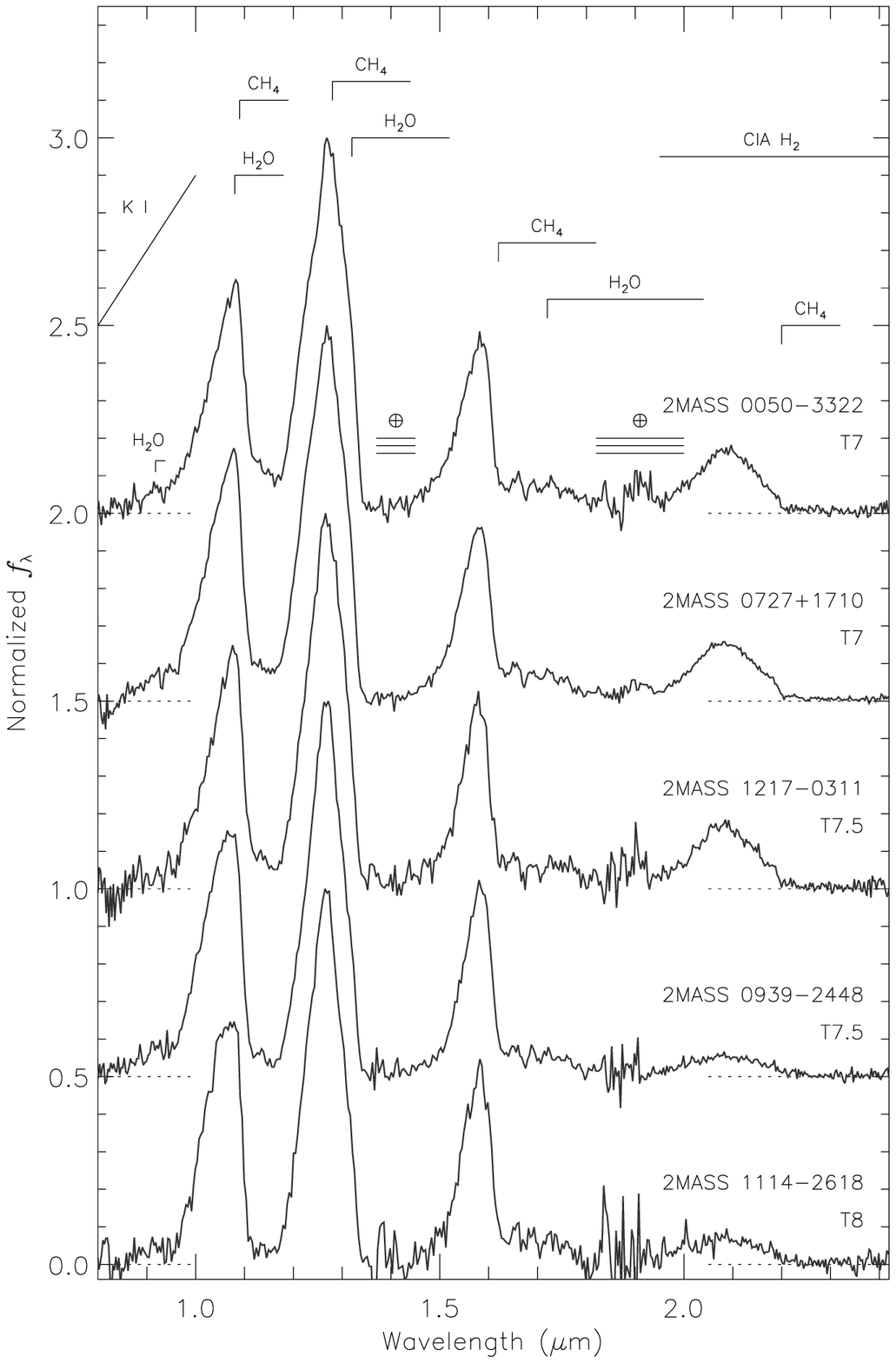}
\caption{SpeX prism spectra for newly observed T dwarfs.
All data are normalized at the 1.27~$\micron$ flux peaks and offset by a constant (dotted lines).
Major spectral features are labelled, and regions of strong telluric absorption are
indicated by $\oplus$ symbols.
\label{fig1}}
\end{figure}

The reduced spectra of the newly observed T dwarfs
are shown in Figure~\ref{fig1}.
Readily apparent are the deep molecular bands of {\water} and {\meth}
that shape the 1.05 ($Y$-band), 1.27 ($J$-band), 1.6 ($H$-band)
and 2.1~$\micron$ ($K$-band) flux peaks, the defining features
of T dwarf near infrared spectra.
The spectra are also shaped by the pressure-broadened red wings of the 0.77 $\micron$
\ion{K}{1} doublet shortward of 1 $\micron$ and collision-induced H$_2$
absorption at $K$-band, both of which are discussed in detail below.
Finer atomic line features,
including the 1.17 and 1.25 $\micron$ \ion{K}{1} doublets, are
unresolved in these data.
Further discussion on the spectral characteristics of T dwarfs
can be found in %\citet{bur00,bur01}; \citet{bur02};
\citet{me02a,me03d};
\citet{geb02}; \citet{mcl03}; \citet{kna04}; \citet{nak04}; \citet{cus05};
and \citet{kir05}.

\subsection{Spectral Signatures of Surface Gravity and Metallicity}

\begin{figure}
\epsscale{1.0}
\plottwo{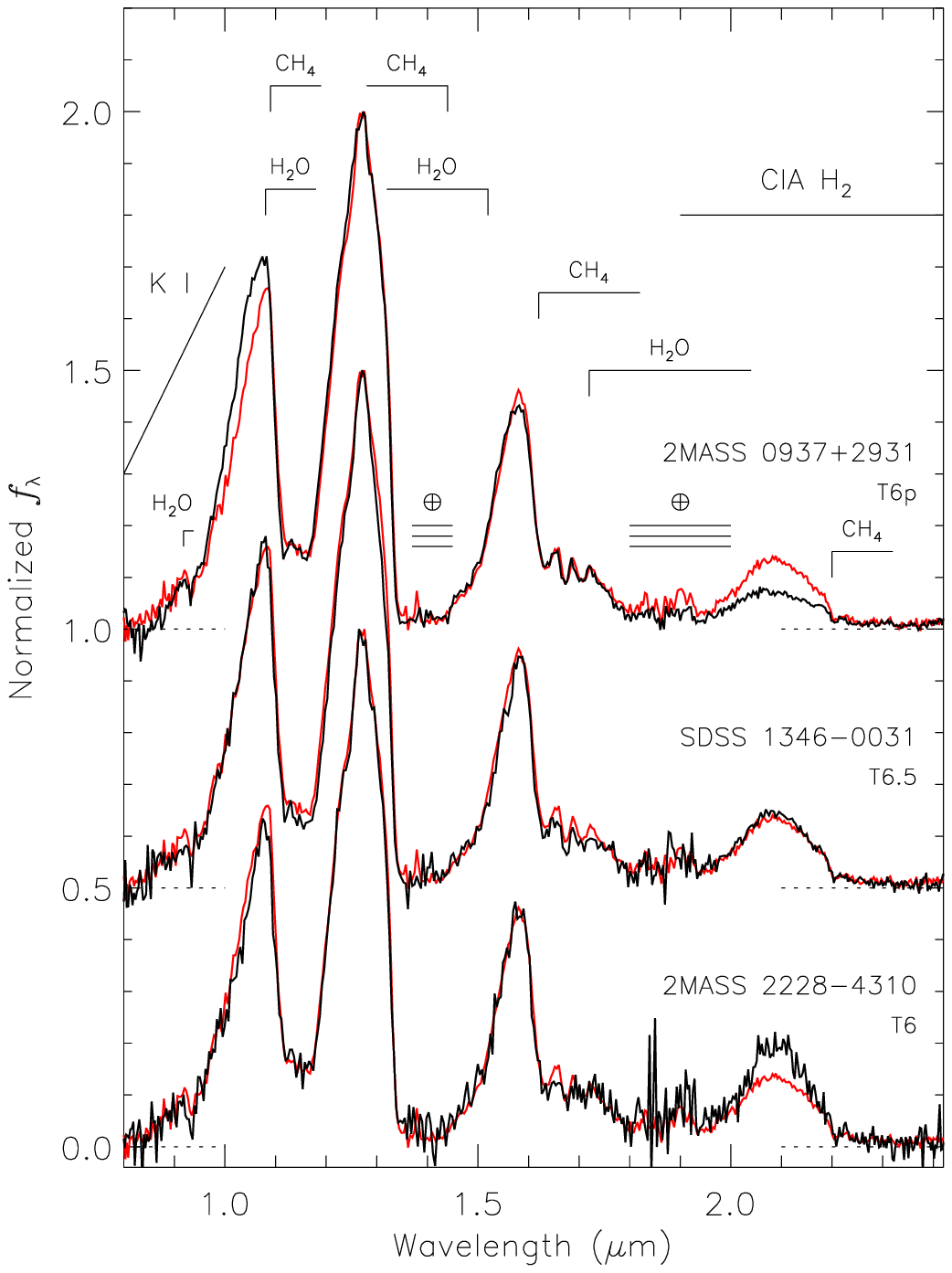}{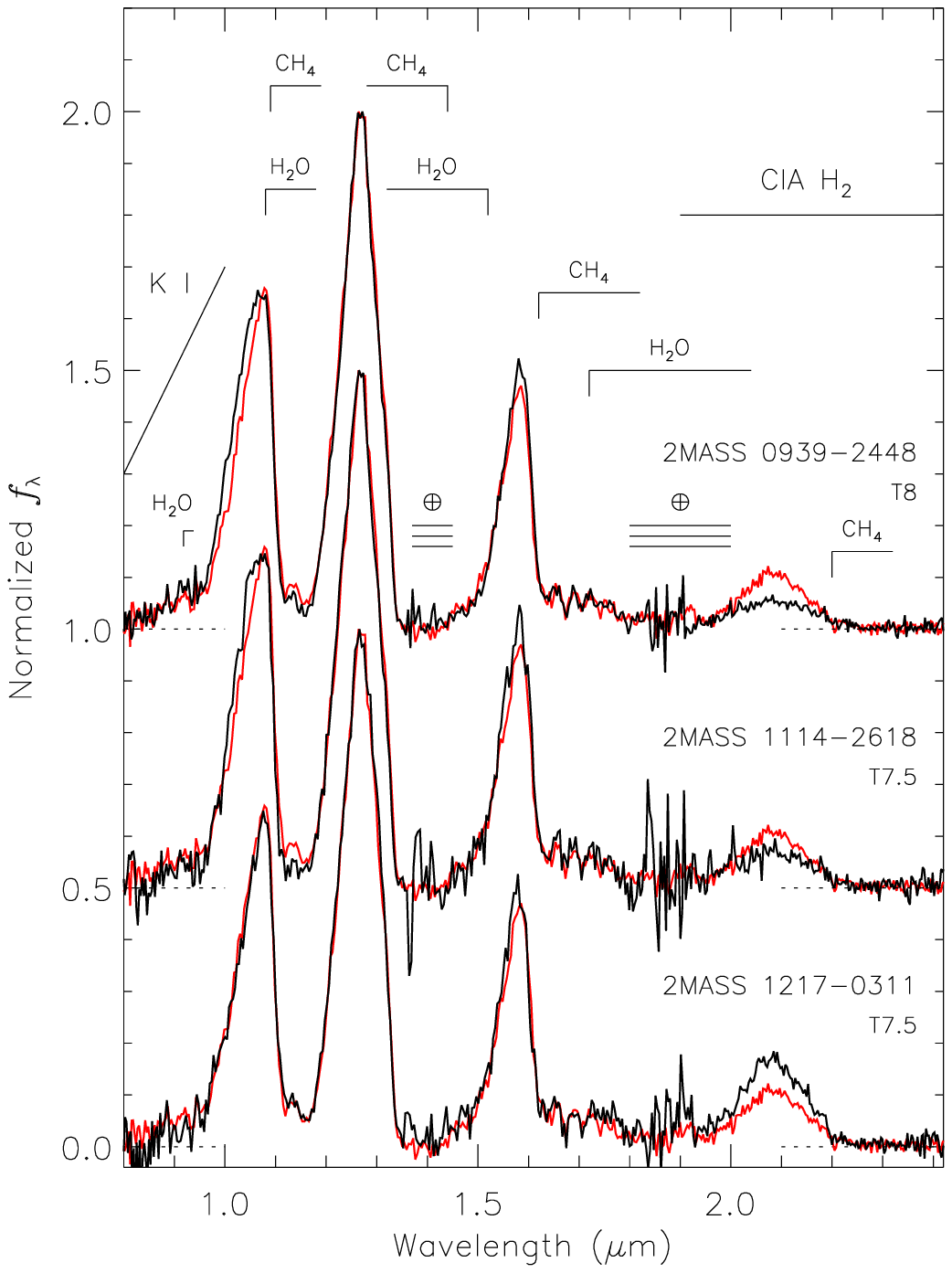}
\caption{Surface gravity and metallicity features in
T dwarf spectra.  The left panel compares the normalized spectra of
the T6-T6.5 dwarfs 2MASS~0937+2931,
SDSS~1346$-$0031 and 2MASS~2228$-$4310
superimposed on that of the T6 spectral standard SDSS~1624+0029 (red line).
The right panel compares the normalized spectra of
the T7.5-T8 dwarfs
2MASS~0939$-$2448, 2MASS~1114$-$2618 and 2MASS~1217$-$0311 with that of
the T7.5 companion brown dwarf Gliese~570D.
Major spectral features are labelled.  Note in particular
the discrepancies at the 1.05 and 2.1 $\micron$ peaks due
to differences in \ion{K}{1} and H$_2$ absorption, respectively.
\label{fig2}}
\end{figure}

Variations in the near infrared spectral features of T dwarfs
are generally synchronized with spectral type -- later subtypes
exhibit both stronger {\water} and {\meth} bands and
bluer near infrared colors.  However,
slight deviations to these trends exist, and are apparent
when one compares sources with similar spectral types,
as in Figure~\ref{fig2}.  Displayed in the left panel of this figure
are the normalized spectra of
three T6/T6.5 dwarfs, 2MASS 0937+2931, SDSS 1346$-$0031 and 2MASS 2228$-$4310,
overlain on that of the T6 spectral standard SDSS~1624+0029 \citep{me05c}.
While {\water} and {\meth} bands are similar among these
spectra, clear differences are seen in the relative brightness
of the $K$-band flux peak and the shape of the $Y$-band peak.
In particular, 2MASS~0937+2931 exhibits weaker $K$-band emission
and a broader $Y$-band flux peak as compared to SDSS~1624+0029,
while 2MASS 2228$-$4310 has stronger $K$-band emission.  Similar
deviations are also seen among the three T7.5/T8 dwarfs
2MASS~0939$-$2448, 2MASS~1114$-$2618 and 2MASS 1217$-$0311 when
compared to the similarly classified Gliese 570D.

What gives rise to these deviations?  Shortward of the $Y$-band spectral peak,
the dominant absorbers in T dwarf spectra are the pressure-broadened wings of the
{\kion} and {\naion} fundamental doublet lines
centered at 0.77 and 0.59~$\micron$, respectively \citep{tsu99,bur00,all03,bur03}.
These features strengthen with later spectral type
throughout the L and T dwarf sequences \citep{kir99,rei00,me03d}.
The broad wings of the alkali lines, induced by kinematic perturbations
by other chemical species
(most importantly H$_2$ and He), are enhanced
in higher pressure ($P$) and higher density atmospheres.
As atmospheric pressure scales
as $dP/d{\tau} \sim P/{\tau} \propto g/{\kappa}_R$ (where $\tau$ is the optical depth and
${\kappa}_R$ the Rosseland mean opacity),
higher pressure photospheres ($\tau = 2/3$) are achieved in brown dwarfs
with higher surface gravities and/or metal-deficient atmospheres (reduced ${\kappa}_R$).
For these sources, line broadening theory \citep{all03,bur03}
predicts the strongest absorption close to the line centers,
resulting in steep 0.8--1.0~$\micron$ spectral slopes due to red wing of the
\ion{K}{1} doublet.

The $K$-band peak, while molded by {\water} and {\meth}
bands at 1.8 and 2.2~$\micron$, is dominated by another pressure-sensitive
feature, collision-induced H$_2$ absorption \citep{lin69,sau94,bor97}.  The induced
1-0 quadrupolar moment of this
molecule produces a broad, featureless absorption centered near 2.1~$\micron$.
Like the {\kion} wings, H$_2$ absorption
arises from kinematic perturbations and is therefore enhanced
in the higher pressure and higher density atmospheres
present on high surface gravity and/or low metallicity brown dwarfs.

While deviations in the strengths of the
{\kion} and H$_2$ features have previously been linked to
gravity and metallicity variations in T dwarfs \citep{me02a,me03d,bur02,leg03,kna04},
Figure~\ref{fig2} demonstrates that these features are correlated.
The steeper {\kion} wings and enhanced H$_2$ absorption exhibited in the spectra of
2MASS~0937+2931, 2MASS~0939$-$2448 and 2MASS~1114$-$2618 are
both indicative
of higher pressure photospheres; while the weaker H$_2$ absorption
in the spectra of 2MASS~2228$-$4310 and
2MASS~1217$-$0311 indicate low pressure photospheres.

In contrast, the congruence of the {\meth} and {\water} bands for similarly
classified T dwarfs suggests that gravity and metallicity effects for these
features are minimal.
The observed correlation between {\teff} and spectral type, the latter based
on the strengths of the molecular bands,
links {\water} and {\meth} to temperature.  However,
gas pressure does regulate the atmospheric abundance of {\meth} and
{\water} in the
principle reaction CO + 3H$_2$ $\rightleftharpoons$ {\meth} + {\water}
\citep{feg96,bur99}, while metallicity modulates both {\meth} and {\water}
abundances \citep{lod02}.  Hence, nearly all of the major absorption features in T dwarf spectra
are affected in some manner by {\teff}, $g$ and [M/H].

\section{Spectral Models}

To further investigate the physical origins of the spectral peculiarities
described above,
we have examined a new suite of brown dwarf spectral models that
incorporate differences in {\teff}, surface gravity and metallicity.  The models,
developed by the Tucson group \citep{bur00,bur02,bur03b,bur05}, are self-consistent,
non-gray atmospheres incorporating up-to-date molecular opacities
as described in \citet{bur01}.  The atmospheres are assumed to be free
of condensate dust species,
consistent with prior modelling results \citep{tsu99,all01},
following the prescription of condensate rainout as described in
\citet[see also Lodders 1999]{bur99}.  Modified Lorentzian profiles
with an ad-hoc, smooth cutoff are
used to model the line broadening of the \ion{Na}{1}
and \ion{K}{1} doublets \citep{bur00}.  Nonequilibrium
mixing effects \citep{sau03} are not considered.
A full description of these models is given in \citet{bur02}.

\begin{figure}
\epsscale{0.6}
\plotone{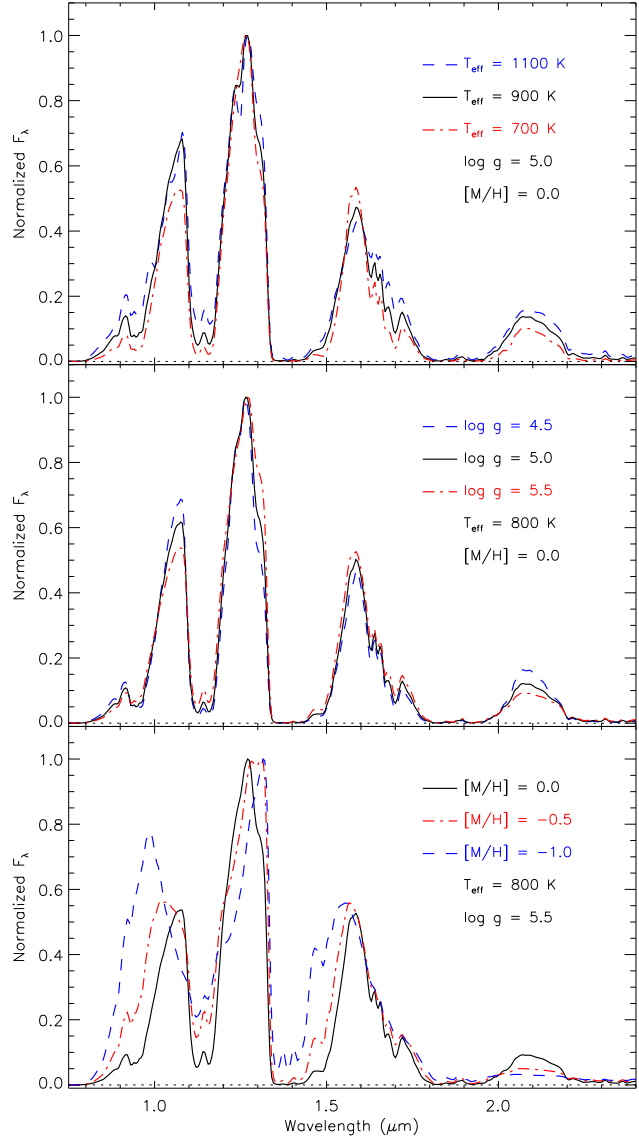}
\caption{Comparison of temperature,
surface gravity and metallicity effects
in T dwarf spectral models.   The top panel displays three
solar metallicity models with $\log{g}$ = 5.0~cm s$^{-2}$ and {\teff} =
1100 (blue dashed line), 900 (black solid line) and 700~K (red dot-dashed
line).  The middle panel displays three solar metallicity models
with {\teff} = 800~K and $\log{g}$ = 4.5 (blue dashed line), 5.0
(black solid line) and 5.5~cm s$^{-2}$ (red dot-dashed line).
The bottom panel displays three {\teff} = 800~K, $\log{g} = 5.5$~cm s$^{-2}$
models with metallicities [M/H] = 0 (black sold line), $-0.5$
(red dot-dashed line) and $-1.0$ (blue dashed line).
Each spectral model has been deconvolved to the
resolution of the SpeX data ({\ldl} $\sim 150$) and normalized
at their $J$-band peak.  \label{fig3}}
\end{figure}

Figure~\ref{fig3} compares three sequences of these spectral models
varying {\teff}, $g$ and [M/H], respectively.  The spectral resolution
of the models
has been degraded using a Gaussian kernel to match
the resolution of the SpeX prism data.  There is general agreement
in the overall spectral morphologies of the models and
observed data; however, important
discrepancies are present.  Most prominent of these
is the shape of the
1.6~$\micron$ CH$_4$ band, reflecting continued
deficiencies in the near infrared opacities of this molecule
for which only low temperature (300 K)
laboratory measurements have been obtained
(Saumon et al.\ 2000; however, see Homeier et al.\ 2003).
These opacities also detrimentally affect
absorption features at 1.1 and 1.3~$\micron$,
although bands at 2.2 and 3.3~$\micron$ have been found
to be adequately reproduced (M.\ Cushing, 2005, priv.\ comm.).
In addition, the line-broadening theory employed by these models predates
more recent calculations by \citet{bur03}, which predict a sharper cutoff
for the red \ion{K}{1} wing at 1.0 $\micron$, in contrast to the modified
Lorentzian profile used here which is relatively flat
over the 0.9--1.0 $\micron$ waveband.  As a result,
alkali opacity at shorter wavelengths in the models is reduced (note the
stronger 0.92 $\micron$ H$_2$O as compared to the data) while the 1.05 $\micron$
$Y$-band peak is more suppressed in the models than observed.

The top panel of Figure~\ref{fig3} shows temperature variations
for solar metallicity and $\log{g} = 5.0$~cm~s$^{-2}$ models.
The trends in this
sequence reflect those observed in T dwarf spectra
as a function of spectral type; i.e., strengthening {\water} and {\meth}
bands producing more acute triangular flux peaks, and stronger
absorption shortward of 1~$\micron$ and at $K$-band.
The increasing depths of the molecular bands with cooler effective temperatures
is largely a consequence of the increased column depth, and therefore
total opacity, of the associated gas species.

The middle panel compares surface gravity variations in the {\teff} = 800 K, solar
metallicity models.  Here, spectral variations are strongest
at the $Y$- and $K$-band flux peaks, although {\water} and {\meth} absorptions
at 1.1 and 1.3 $\micron$, and the $H$-band peak, are also affected.
The $K$-band peak is suppressed in
higher surface gravity models, consistent with the
enhanced H$_2$ absorption expected in higher pressure photospheres.
On the other hand, surface gravity variations at the
$Y$-band peak do not match the observed trends.
The highest surface gravity model exhibits reduced 1.05~$\micron$ flux
relative to 1.27~$\micron$, contrary to the brighter and
broadened $Y$-band peaks seen in the empirical data.
We attribute this discrepancy to the outdated alkali line
broadening profile employed in these models, and leave analysis
of this feature to a future study.

\begin{figure}
\epsscale{1.0}
\plottwo{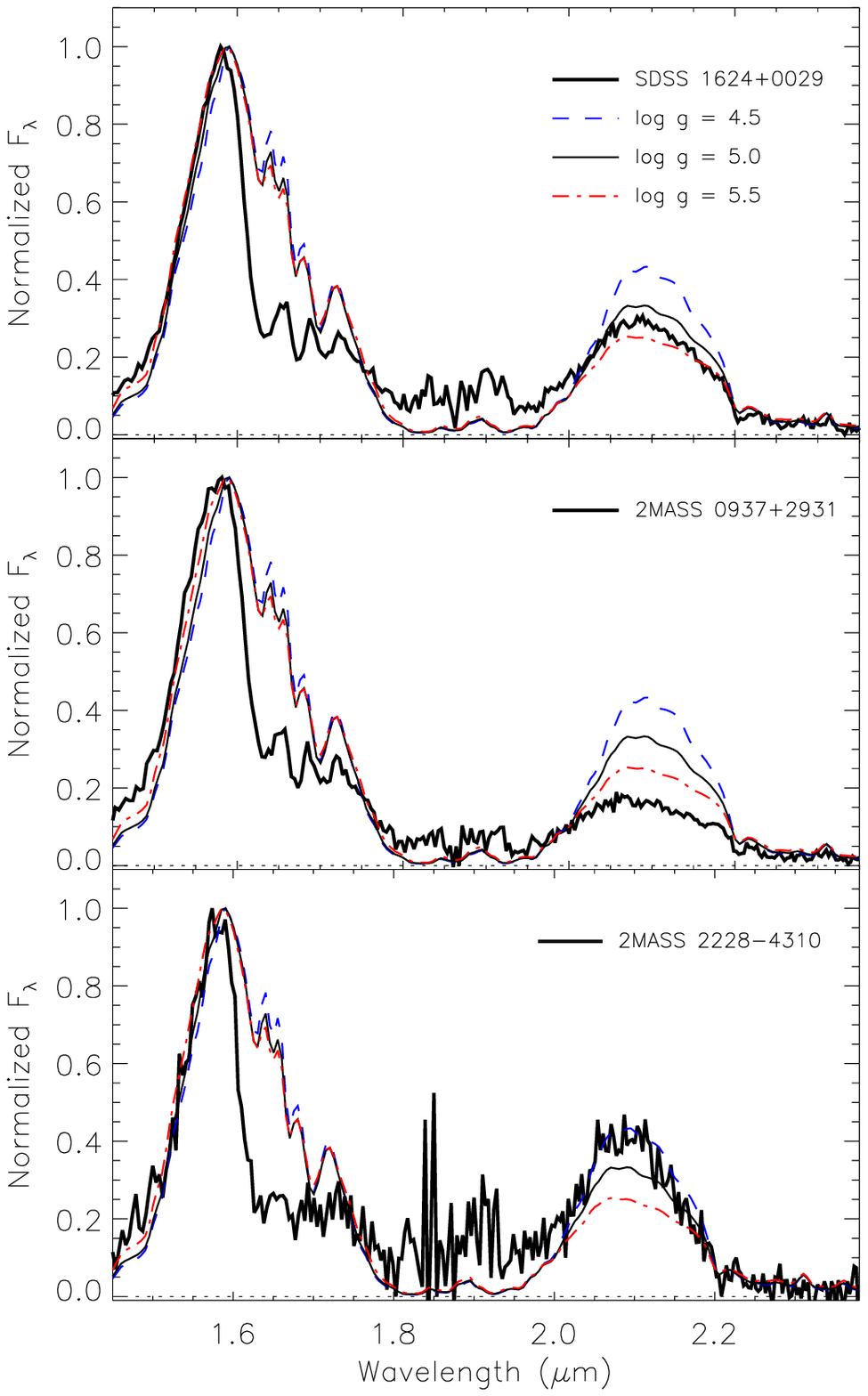}{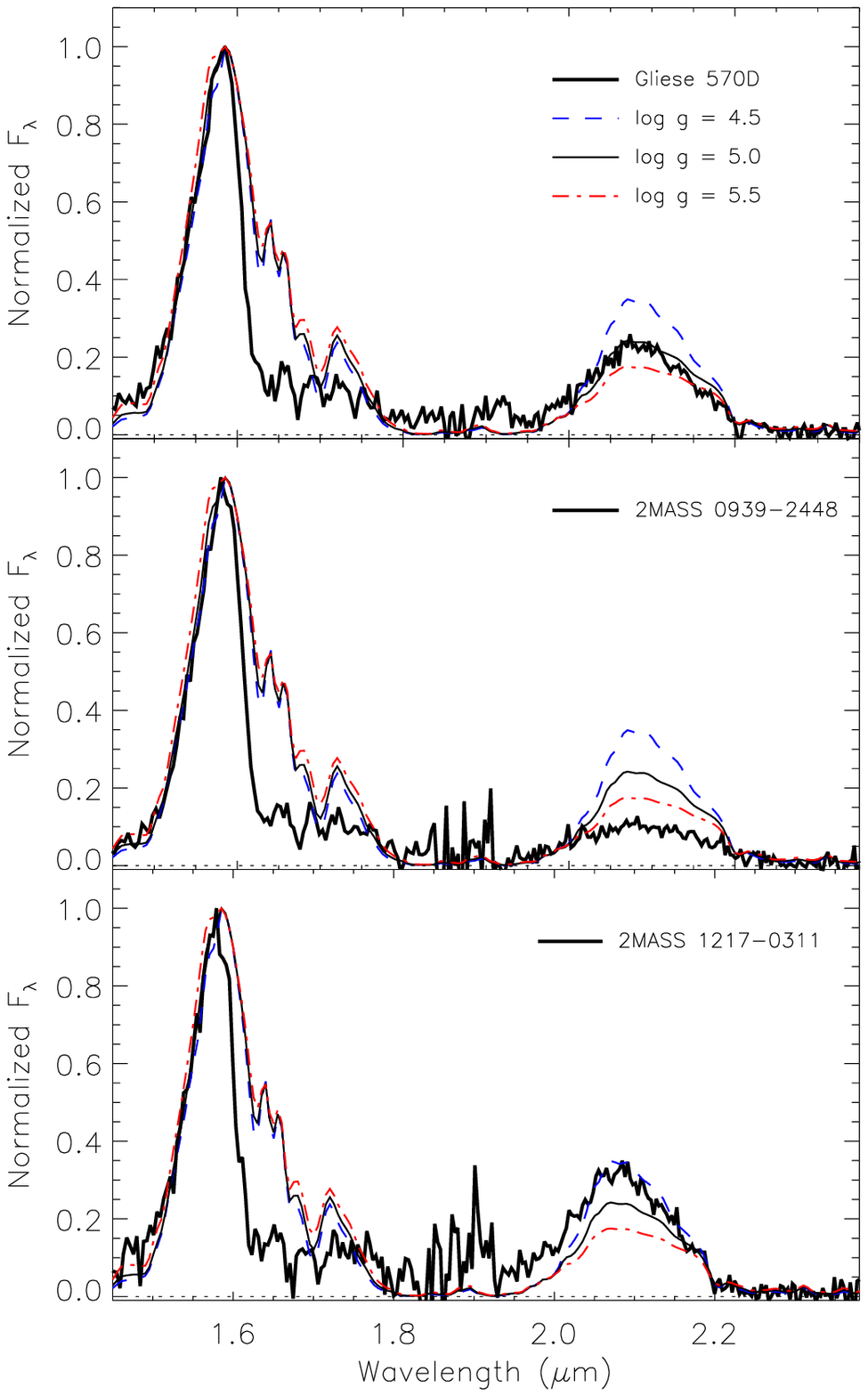}
\caption{Comparison of models to spectra in the 1.5--2.4 $\micron$
spectral region.  Shown are data (solid black lines) for the T6-T6.5 dwarfs
SDSS~1624+0029, 2MASS~0937+2931 and 2MASS~2228$-$4310 (left panel)
and the T7.5-T8 dwarfs Gliese~570D, 2MASS~0939$-$2448 and 2MASS~1217$-$0311
(right panel).  Resampled solar metallicity spectral models for $\log{g}$ = 4.5
(blue dashed line), 5.0 (black solid line) and 5.5~cm s$^{-2}$ (red dot-dashed line)
and {\teff} = 800~K (left panel) and 1000~K (right panel) are overplotted.
Both data and models are normalized at the $H$-band peak. \label{fig4}}
\end{figure}

The similarity of surface gravity modulations of the
$K$-band flux peak in the theoretical models
to variations in the spectral data shown in Figure~\ref{fig2}
is highlighted in Figure~\ref{fig4}, which
shows a similar sequence of T6/T6.5 and T7.5/T8 dwarfs
but overlain on the {\teff} = 1000 and 800 K solar metallicity models,
respectively.  Inequities in
the 1.6~$\micron$ {\meth} band notwithstanding, the spectral
models reproduce reasonably well the relative variations observed in the
$K$-band flux peaks, although both
2MASS~0937+2931 and 2MASS~0939-2448 exhibit stronger $K$-band
suppression than the highest surface gravities permitted
by the models.

The bottom panel of Figure~\ref{fig3}
compares metallicity effects at fixed surface gravity
($\log{g} = 5.5$) and temperature ({\teff} = 800 K)
for [M/H] = 0, $-0.5$ and $-1$.
The higher surface gravity examined here is appropriate for old brown dwarf
members of the metal-poor Galactic thick disk and halo populations.
Spectral variations are far more extreme in this case.
The $K$-band peak is suppressed at lower metallicities,
as expected for enhanced H$_2$ absorption.  At the same time,
emergent flux appears to be relatively enhanced shortward of 1~$\micron$,
due to reduced Na and K
abundances and their corresponding opacities.  The specific shape of the spectrum
at these wavelengths should be
treated with caution, however, given the older line broadening theory used in the models.
Shifts in the $J$- and $H$-band peak wavelengths are due to
reduced H$_2$O abundances
and increased collision-induced H$_2$ absorption extending into the $H$-band.
Overall, these spectral variations are more substantial
than those seen in the empirical data, a sign
that significant subsolar metallicities ([M/H] $\lesssim -0.5$)
are not present among the T dwarfs examined here.
However, the broadened $Y$-band peaks and strong $K$-band
suppression in the spectra
of 2MASS~0937+2931, 2MASS~0939$-$2448 and 2MASS~1114$-$2618
do hint at slightly subsolar metallicities for these sources.

\section{Measuring Physical Properties from T Dwarf Spectra}

\subsection{The Method}

The spectral models confirm that the
pressure-sensitive H$_2$ and \ion{K}{1} absorptions
are more strongly influenced by surface gravity effects than the absorption bands
of {\water} and {\meth}, while the latter vary more strongly with {\teff},
at least for the temperature regime considered here.
By contrasting the strengths of these features, it should
be possible in principle to extract the
effective temperatures and surface gravities of these objects.
%by comparing their
%low resolution spectra to theoretical models.
In practice, this pursuit has proven problematic due to persistent
inadequacies in molecular opacities and corresponding systematic errors
in the models (e.g., Burgasser et al.\ 2004a).  What is required is a means
of calibrating the spectral models using one or more empirical fiducials.

Fortunately, such a fiducial exists in the T dwarf Gliese 570D \citep{me00a}.
This widely-separated ($\rho$ = 258$\arcsec$ = 1530 AU),
common proper motion brown dwarf companion
to the nearby (5.91$\pm$0.06 pc; ESA 1997) Gliese~570ABC system
has both distance and luminosity measurements that are empirically well constrained.
The age of this system is estimated to be 2--5~Gyr
based on a comparison of age, activity and kinematic relations for the K and M stellar
components \citep{me00a,geb01}. The K4~V primary has a near-solar
metallicity ([Fe/H] = 0.00--0.16; Feltzing \& Gustafsson 1998;
Thoren \& Feltzing 2000; Allende Prieto \& Lambert 2000).
Assuming coevality, Gliese~570D is one of the few T dwarfs
with both age and metallicity constraints.
\citet{geb01} derive fairly precise
temperature ({\teff} = 784--824 K) and
surface gravity ($g =$ (1--2)${\times}10^5$ cm s$^{-2}$)
estimates, the former based on the measured luminosity and a model-dependent
radius, the latter based on
brown dwarf evolutionary models \citep{bur97}.

Using Gliese~570D as our empirical fiducial,
our procedure was then as follows.  The strengths of the major {\water}
bands and relative fluxes of the spectral peaks were measured
for both the empirical data sample and theoretical models
using the following ratios:
\begin{equation}
H_2O-J = \frac{\int{F_{1.14-1.165}}}{\int{F_{1.26-1.285}}},
\end{equation}
\begin{equation}
H_2O-H = \frac{\int{F_{1.48-1.52}}}{\int{F_{1.56-1.60}}},
\end{equation}
\begin{equation}
Y/J = \frac{\int{F_{1.005-1.045}}}{\int{F_{1.25-1.29}}},
\end{equation}
\begin{equation}
K/J = \frac{\int{F_{2.06-2.10}}}{\int{F_{1.25-1.29}}}
\end{equation}
and
\begin{equation}
K/H = \frac{\int{F_{2.06-2.10}}}{\int{F_{1.56-1.60}}},
\end{equation}
where $\int{F_{{\lambda}_1-{\lambda}_2}}$ denotes the integrated flux
between wavelengths ${\lambda}_1$ and ${\lambda}_2$ (in microns).
The first two ratios are identical to those defined for the
near infrared classification of T dwarfs; the spectral region
sampled by these is shown in Figure~5 of \citet[note that
ratios sampling the poorly modelled {\meth} bands are not considered here]{me05c}.
The $K/J$ ratio has also been used previously to examine variations in H$_2$ absorption
\citep{me04b,me05c}.  The $Y/J$ and $K/H$ color ratios are defined here for
first time.
%(the spectral ratios sampled by these are indicated in Figure~4).
Measurements of these ratios
on the empirical data are given in Table~\ref{tab:ratios}.

\begin{figure}
\epsscale{0.9}
\plottwo{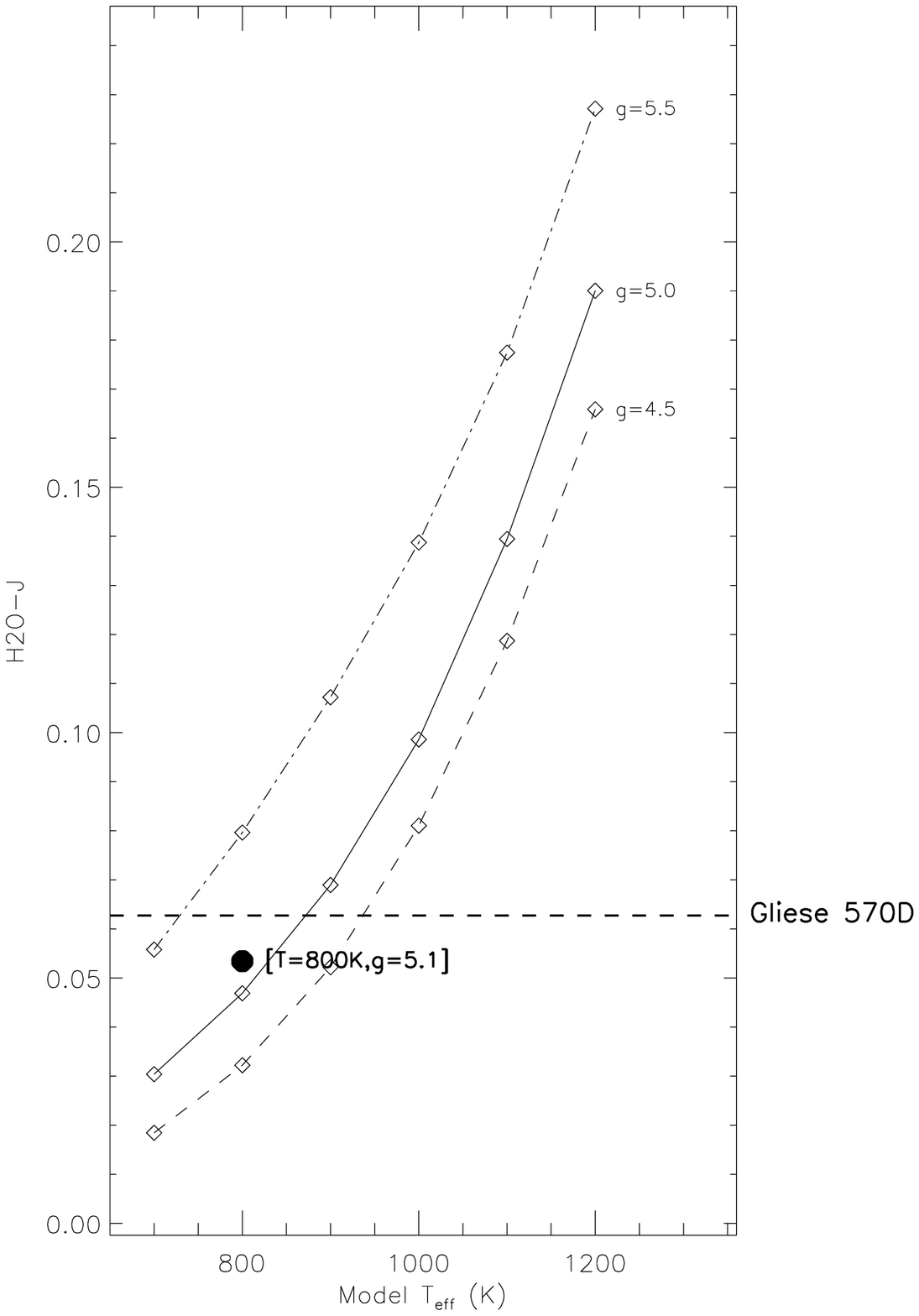}{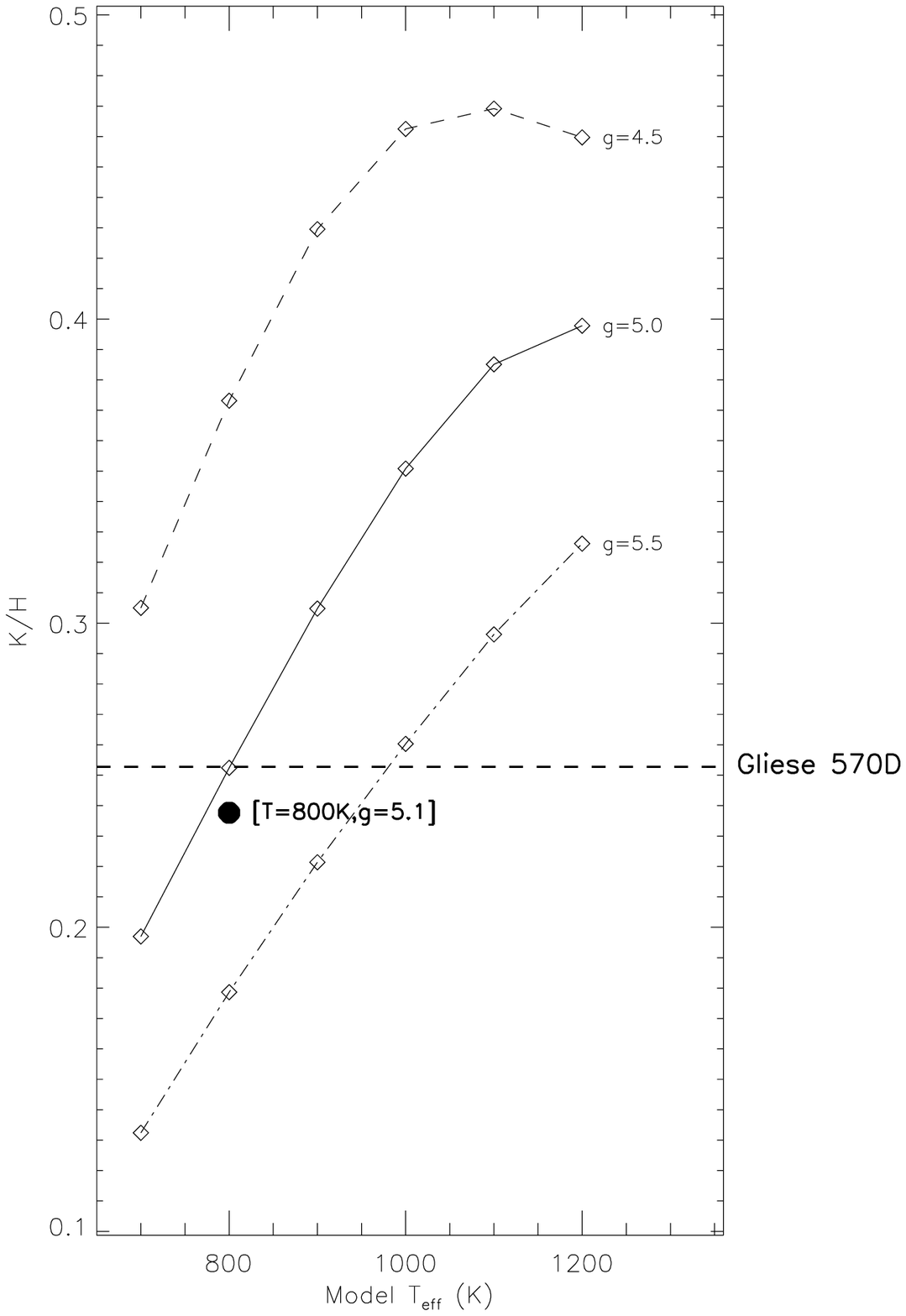}
\caption{Values for the spectral ratios {\water}$-J$
and $K/H$ as measured on solar metallicity models
(points with dash-dot lines connecting constant surface
gravity models).  Values measured on the spectrum of
Gliese~570D are indicated by the dashed lines.
The solid circle indicates the
interpolated model index value for the adopted
physical parameters
of Gliese~570D, {\teff} = 800~K, $\log{g}$ = 5.1~cm s$^{-2}$
and [M/H] = 0.
\label{fig5}}
\end{figure}

A series of {\water}$-J$
and $K/H$ ratios for the solar metallicity models
are shown in Figure~\ref{fig5}.  Both ratios vary according to
differences in {\teff}
and $\log{g}$, although gravity variations
are stronger in the $K/H$ color ratio.  Gravity
variations also affect the two ratios in opposite ways; higher
gravity models yield large {\water}$-J$ values (implying
weaker absorption) and smaller $K/H$ values (implying weaker
$K$-band emission).  Similar trends are seen with the {\water}$-H$ and
$K/J$ ratios, respectively.  All of these trends are consistent with the qualitative
properties of the model spectra shown in Figure~\ref{fig3}.

Calibration of the model ratios was achieved by correcting these values
to those measured from the SpeX spectrum of Gliese 570D.  Adopting {\teff} = 800 K and
$\log{g}$ = 5.1 for this source, we
computed the corresponding model ratios by linear interpolation.
Correction factors, listed at the bottom of Table~\ref{tab:ratios},
were defined as the ratio of the
spectral data measurement to the model value.
For four of the ratios,
model values differ by less than 20\%; the {\water}$-H$ index, on the other hand, requires
a 60\% correction.  The correction factors
were applied to all of the solar metallicity
model ratios, regardless of {\teff} or gravity,
and therefore represents a first order calibration of the models.

In principle, any of the
ratios defined in Eqns.~1--5 could be used for comparison to the spectral data.
We restrict our primary analysis
to the {\water}$-J$ index, which requires
a smaller correction factor than {\water}$-H$; and
the $K/H$ index, which gives a quantitative measure of the behavior demonstrated
in Figure~\ref{fig4}. The corrected model ratios were
resampled in steps of 20 K in {\teff} and 0.1 dex in $\log{g}$
by linear interpolation. Then, for each spectrum, we identified
the region in {\teff} and $\log{g}$
phase space for which the corrected model ratios agreed with the empirical
ratios, assuming a 10\% uncertainty (see $\S$~4.3).

\begin{figure}
\epsscale{0.8}
\plotone{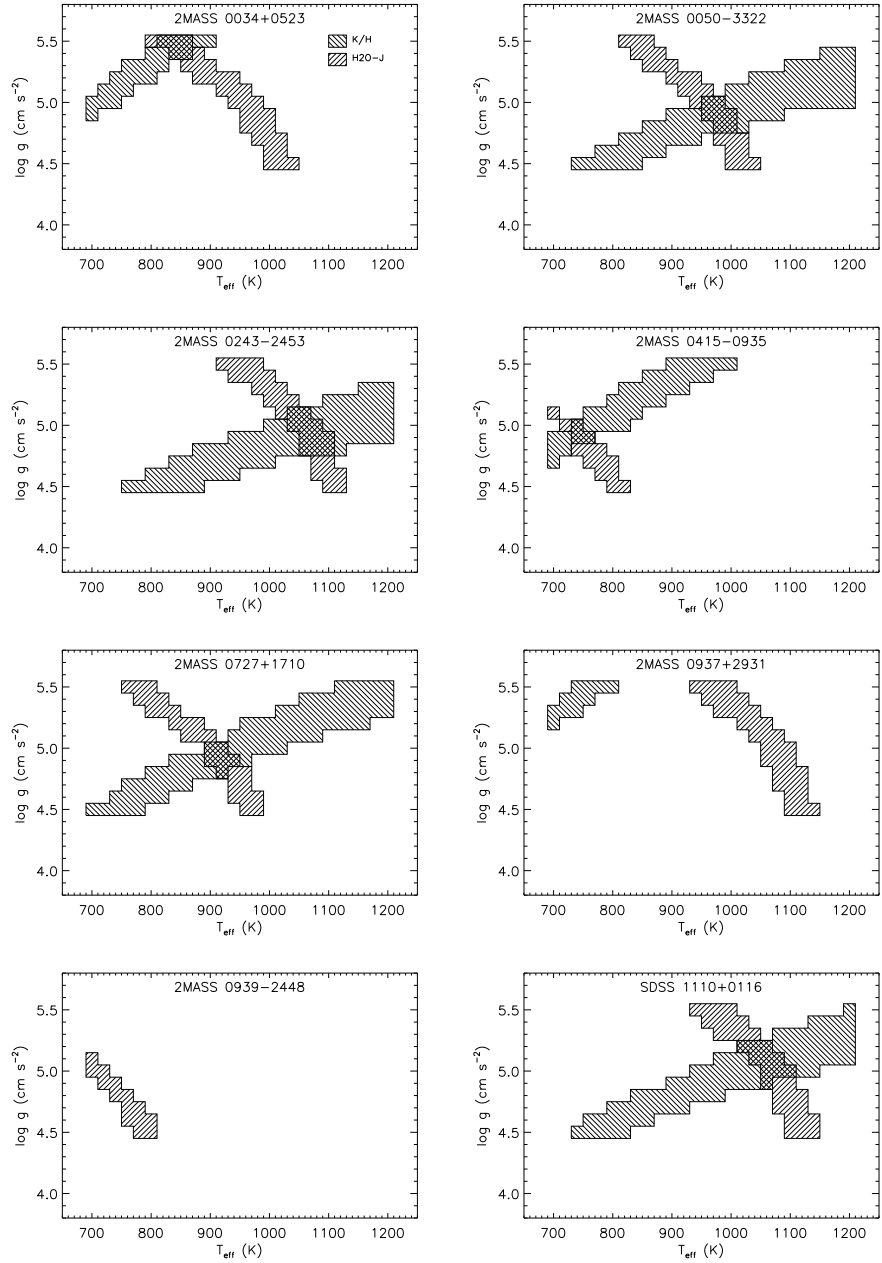}
\caption{Derived physical parameter phase spaces for
the T dwarfs 2MASS~0034+0523, 2MASS~0050$-$3322, 2MASS~0243$-$2453,
2MASS~0415$-$0935, 2MASS~0727+1710, 2MASS 0937+2931,
2MASS~0939$-$2448 and SDSS 1110+0116.
{\teff} and $\log{g}$ values for which measurements
of the spectral ratios {\water}$-J$ and $K/H$
match scaled values for the models (assuming an uncertainty of 10\%)
are indicated by hatched regions.
The overlap cross-hatched regions represent our best match
``fits'' for these parameters. \label{fig6}}
\end{figure}

\begin{figure}
\epsscale{0.9}
\plotone{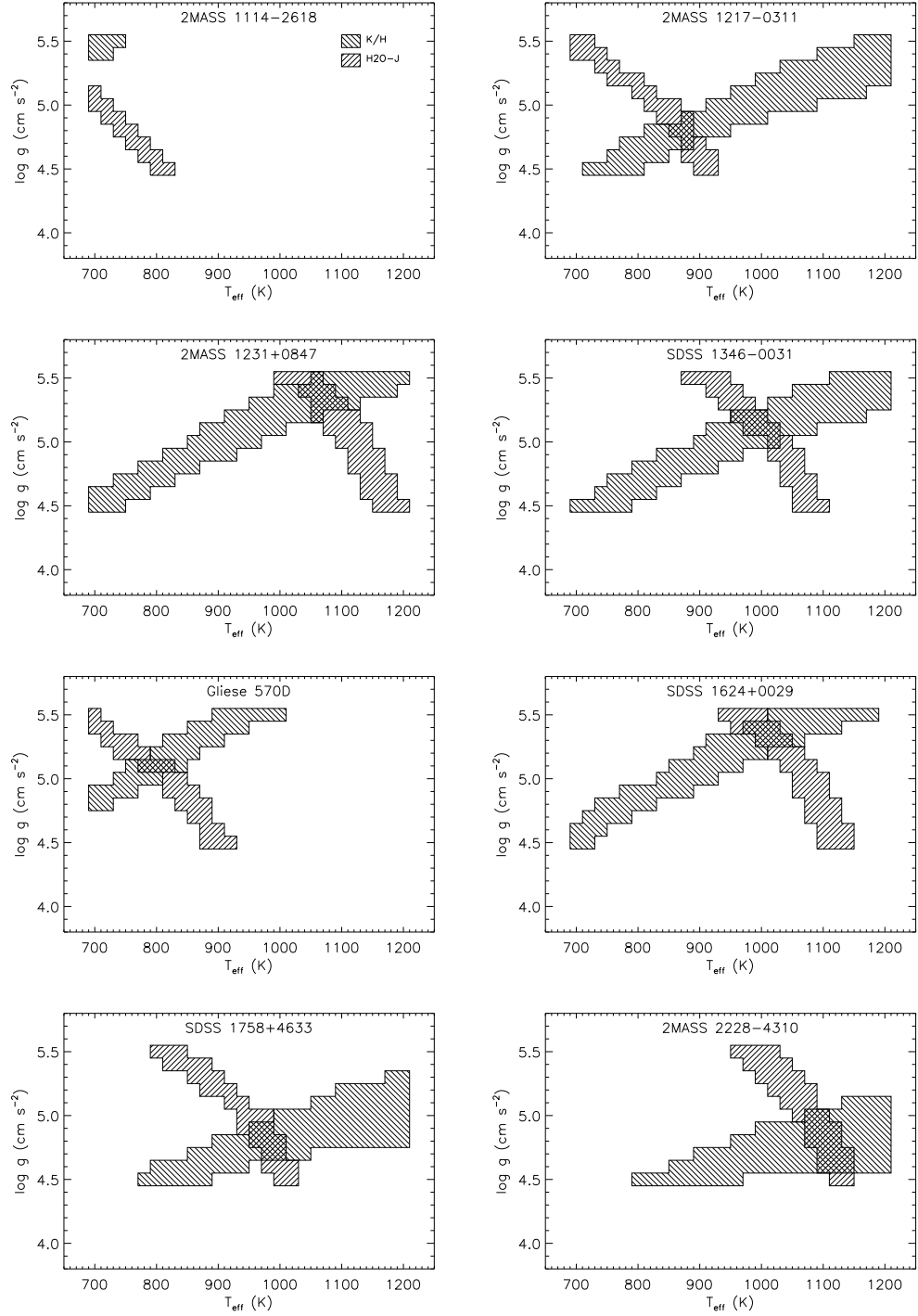}
\caption{Same as Figure~\ref{fig6} for the T dwarfs
2MASS~1114$-$2618, 2MASS~1217$-$0311, 2MASS~1231+0847,
SDSS~1346$-$0031, Gliese~570D, SDSS~1624+0029, SDSS~1758+4633
and 2MASS~2228$-$4310. \label{fig7}}
\end{figure}

Figures~\ref{fig6} and~\ref{fig7} illustrate these matches.
The {\water}$-J$ and $K/H$ ratios each constrain a set
of {\teff} and $\log{g}$ values that span the model
parameter space diagonally; e.g., agreement in the
{\water}$-J$ ratios span
low temperatures and high surface gravities to high temperatures
and low surface gravities.  The $K/H$ ratios match an orthogonal
phase space.  The intersection of these phase spaces
provides an unambiguous constraint on both {\teff} and $\log{g}$.

\subsection{Results}

\subsubsection{{\teff} and $g$ Estimates}

Table~\ref{tab:tg} lists the ranges of
{\teff} and $\log{g}$ constrained by the two ratios for each source in
our sample.  For 13 of the 16 sources (including Gliese~570D), these values are well defined,
with typical uncertainties of 40--60~K in {\teff} and 0.1--0.3~dex in
$\log{g}$.  For three of the T dwarfs, 2MASS~0937+2931, 2MASS~0939$-$2448
and 2MASS~1114$-$2618, no phase space intersection was found.  The case of 2MASS~0937+2931
is discussed in further detail below.  For 2MASS~0939$-$2448
and 2MASS~1114$-$2618, close examination of Figures~\ref{fig6}
and~\ref{fig7} suggests that phase space intersections are possible at lower {\teff}s
than those spanned by our model set; i.e., for {\teff} $\lesssim$ 700~K.  This is intriguing,
since we derive {\teff} = 740--760~K for the T8 2MASS~0415$-$0935,
currently the coldest and lowest luminosity brown dwarf
known \citep{me02a,gol04,vrb04}.  Parallax measurements can
determine whether
2MASS~0939$-$2448 and 2MASS~1114$-$2618 are in fact colder and fainter
brown dwarfs.

Are these temperatures and surface gravities reasonable?  Eight of the T dwarfs in our
sample have prior {\teff} determinations from \citet{gol04} and \citet{vrb04}
based on parallax and bolometric luminosity ($L$) measurements; these values are
listed in Table~\ref{tab:tg}.  In all eight cases, our derived {\teff}s are
consistent.  This agreement may have much to do with
the large {\teff} estimate ranges from \citet{gol04} and \citet{vrb04},
as high as 300~K, due to uncertainties in the radii adopted
to compute {\teff} from $L$.  Our {\teff} estimates are typically in the
middle or high end of the ranges from these studies. The only surface gravity
estimates for field T dwarfs reported to date are those of \citet{kna04},
based on a comparison of near infrared colors to atmosphere models by
\citet{mar02}.  In this case, we find that our estimates
are systematically 0.3--0.5 dex higher than the Knapp et al.\ values.  As the latter
are stated without uncertainties, we simply point out this discrepancy
for further study.

\subsubsection{A Subsolar Metallicity for 2MASS~0937+2931}

The parameter spaces for 2MASS~0937+2931 do not intersect
in Figure~\ref{fig6}, but it appears that they would if
higher surface gravities were modelled.  However,
surface gravities larger than $\log{g} = 5.5$
are restricted by the interior physics \citep{bur97}.
An alternate hypothesis is that the spectrum is influenced by
a third parameter, namely metallicity.

\begin{figure}
\epsscale{0.9}
\plotone{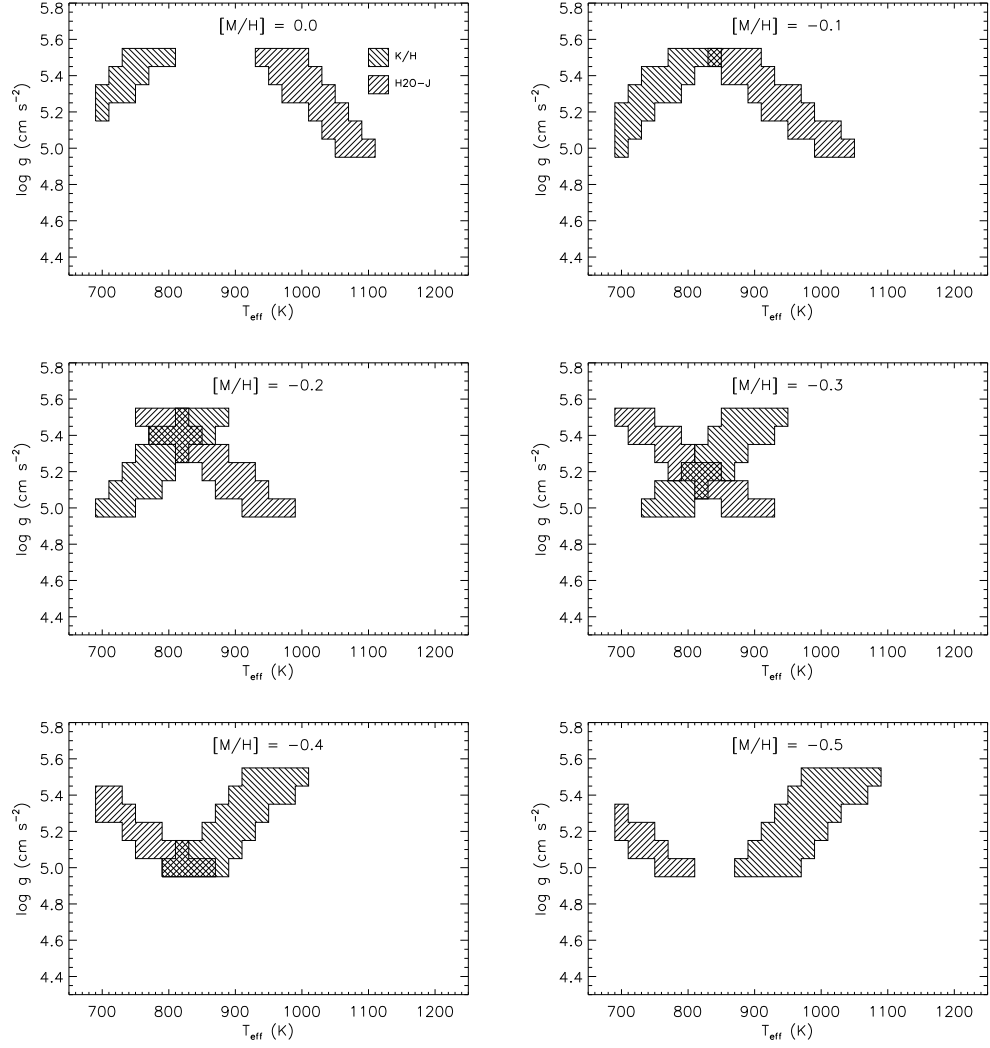}
\caption{{\teff} and $\log{g}$ parameter fit for 2MASS~0937+2931
for spectral models with metallicities $-0.5 \leq$ [M/H] $\leq 0$,
interpolated in steps of 0.1 dex.
\label{fig8}}
\end{figure}

We can quantitatively test this case
by introducing metallicity variations into the model
set.  Applying the same corrections to the model ratios
as above (i.e., assuming Gliese 570D has [M/H] = 0),
linearly interpolating between the [M/H] = 0 and [M/H] = $-0.5$
models in 0.1 dex steps
for 700 $\leq$ {\teff} $\leq$ 1200 K and 5.0 $\leq$ $\log{g}$ $\leq$ 5.5,
and performing the same comparative analysis,
we derive the series of parameter phase spaces (effectively slices of a
three dimensional parameter phase volume)
shown in Figure~\ref{fig8}.
We find that, for the case of 2MASS~0937+2931, the phase spaces
intersect for metallicities $-0.1 \leq$ [M/H] $\leq -0.4$,
780 $\leq$ {\teff} $\leq$ 860 and
5.0 $\leq$ $\log{g}$ $\leq$ 5.5.  Lower metallicities
may also be feasible at lower surface gravities.
One other source in our sample,
2MASS~0034+0523,
also exhibits intersecting parameter spaces for slightly
subsolar metallicities, $-0.2 \leq$ [M/H] $\leq 0$.
This object has also been noted for its strong $K$-band suppression \citep{me04b}.

Our analysis represents the first quantitative
constraints of metallicity for a brown dwarf, and are consistent
with prior qualitative conclusions.
However, the derived values should be treated with
caution for reasons other than the fidelity of the theoretical models.
Constraining the three
parameters {\teff}, $g$ and [M/H] cannot be done unambiguously
using only two spectral ratios; at least one
additional constraint is required.  A promising candidate
is the $Y/J$ ratio, as the metal-poor models shown
in Figure~\ref{fig5} exhibit significant variations
at these short wavelengths.  We defer more thorough
examination of this third index
until such time as the current generation of spectral models incorporate a
more rigorous line broadening theory.

\subsection{Assessing Systematic Effects}

While the {\teff} and $\log{g}$ estimates made here are
reasonable and generally consistent with prior work,
it is important to identify and characterize any source
of systematic error that may skew the results.
Such systematic effects can arise from the calibration or
quality of the spectral data, the choice and calibration
of the spectral ratios used and limitations of the
spectral models themselves.  We examine these effects here in detail.

\subsubsection{Flux Calibration of the Spectral Data}

Accurate measurement of color ratios such as $K/H$ assumes
that the spectral data
portrays the true color of the source, which brings into
question possible reddening of the observed spectrum
and the accuracy of the overall flux calibration.
Interstellar reddening can generally be ruled out for our sample
as all of the
sources lie at distances of $\sim$20 pc or less.
A more local source of reddening, differential color refraction through our atmosphere,
has been mitigated by observing the sources with the slit aligned at
the parallactic angle.  We therefore assume that both of these effects are negligible.

Systematic errors incurred in the flux calibration can be quantified by comparing
spectrophotometric colors for the data to published photometry.
We examined $J-H$, $H-K$ and $J-K$ colors
on the Mauna Kea Observatory system (MKO; Simons \& Tokunaga 2002; Tokunaga, Simons
\& Vacca 2002) using photometry from \citet{geb01,leg02}; and \citet{kna04};
and $J-H$, $H-K_s$ and $J-K_s$ colors
from 2MASS (Table~\ref{tab:sample}).
Spectrophotometric colors were determined by integrating the appropriate
filter profile (combined with telescope and instrumental optical response
curves for 2MASS photometry; see $\S$~III.1.b.i in Cutri et al.\ 2003)
over the near infrared spectra of each T dwarf and that of the A0 V star Vega
\citep[see also Stephens \& Leggett 2003]{ber95}.  We found no systematic differences
for any of the photometric and spectrophotometric colors on both systems,
and typical deviations were 5\% or less for the more accurate MKO photometry.
For those few sources with color offsets significantly greater than their
photometric uncertainties (2MASS~0243$-$2453, 2MASS~0727+1710 and SDSS~1758+4633
have 3$\sigma$ deviations in MKO $J-K$), differences were at most 15\%.  Hence,
we conclude that the 10\% uncertainties
adopted for the color ratios adequately compensates
for uncertainties in the flux calibration.

\subsubsection{Spectral Noise}

Molecular band ratios are generally insensitive to color errors in
the overall spectrum, but deep absorption bands can be affected by spectral noise.
This is manifested by variations in the measured flux at the bottom
of the bands where signal-to-noise (S/N) is minimal.
To explore the impact of this effect, we performed a Monte Carlo experiment,
measuring the {\water}$-J$
ratio on the combined spectrum of Gliese~570D plus a Gaussian
noise component scaled to S/N = 10--200 at the $J$-band peak.
A total of 10~000 trials were run over a uniform range of S/N.
We found no systematic deviations in the {\water}$-J$ ratios for S/N $\gtrsim$ 10,
but scatter among the values increased in the noisier spectra,
approaching 10\% for S/N = 40.  All of our spectra have
S/N $\gtrsim$ 50 at the $J$-band peak with the exception of SDSS~1110+0116, which
has S/N $\approx$ 20.  Hence, we find that
our adopted 10\% uncertainties for the measured {\water} ratios
incorporate noise effects sufficiently, although derived values for SDSS~1110+0116
may be more uncertain.

\subsubsection{Choice of Spectral ratios}

The use of the {\water}$-J$ and $K/H$ ratios in our analysis above
was justified by the sensitivity of these ratios to {\teff} and $g$
variations in the models, the magnitude of the calibration correction
required and the fidelity of the models in these spectral regions.
However, we can also consider how the results change
if a different set of ratios are employed.
We repeated our analysis with four pairings among the
{\water}$-J$, {\water}$-H$, $K/H$ and $K/J$ ratios.
Again, we found no significant or systematic differences
in the derived
{\teff} and $\log{g}$ values amongst our sample, although
parameters for individual sources differed by as much as
160~K and 0.6~dex, respectively.  Typical deviations were
of order 65~K in {\teff} and 0.15~dex in $\log{g}$, which we
adopt as estimates of systematic uncertainty.

\subsubsection{Calibration of the Spectral ratios}

The calibration of the spectral ratios
hinges largely on the assumed physical properties for our calibrator
source Gliese~570D.  But how sensitive are the results to these
adopted parameters?
By varying the assumed {\teff} and $\log{g}$ for Gliese~570D by $\pm$25~K
and $\pm$0.1 dex, respectively, consistent with the range of values found
by \citet{geb01}, we found mean offsets of $\pm$35~K and $\pm$0.1 dex
in the derived parameters for the field sources.  These
offsets were independent; changing the adopted {\teff} of Gliese~570D
had no impact on the derived surface gravities, and vice versa.  Hence, we
estimate that additional systematic
uncertainties of 35~K and 0.1~dex in
{\teff} and $\log{g}$, respectively, accommodate uncertainties
in the physical parameters of our comparison source.

\subsubsection{Choice of Spectral Models}

Many of the potential systematic effects involving the choice and calibration
of the spectral ratios would be eliminated if the models accurately
reproduced the observed spectra.  As limitations in the
opacities prevent this, we must also consider how dependent
our results are on the choice of spectral models used.  We therefore
repeated our analysis using the COND models of \citet{all01}.
Calibration of the spectral ratios was performed in
the same manner as with the Tucson models, yielding somewhat different
correction values.  While deviations typically of order 50~K and 0.1~dex
were found when comparing derived {\teff} and $\log{g}$ between the model
sets, these deviations were not systematic.  Hence, we conclude that the choice
of spectral model does not systematically change our results.

In summary, we find no systematic deviations in our method that would lead
to skewed estimates of {\teff} and $\log{g}$, although systematic
uncertainties of order 50--100~K and 0.1--0.25~dex may be present.

\section{Mass, Radius and Age Estimates}

\subsection{Evolutionary Models}

\begin{figure}
\epsscale{0.9}
\plotone{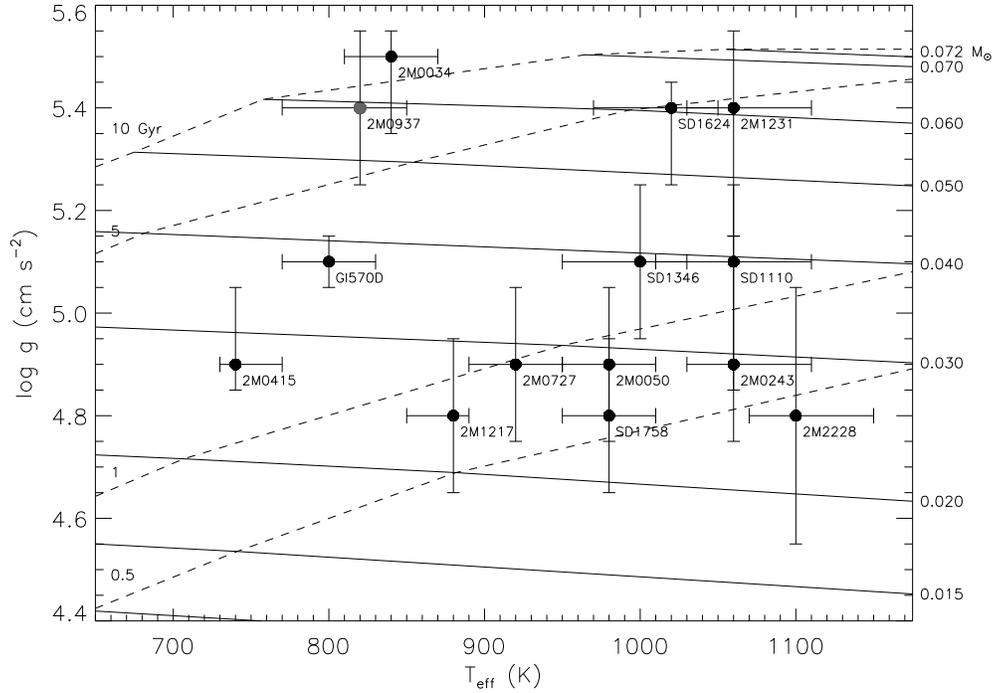}
\caption{Best-fit {\teff} and $\log{g}$ values derived for the T dwarfs
as compared to evolutionary models from \citet{bur97}.  Isochrones
are indicated by dashed lines (10, 5, 2 and 1 Gyr from top to bottom),
while constant mass values (labelled, in Solar masses) are denoted by solid lines.
Error bars on the data points are based on the breadth of the {\teff}, $\log{g}$
space spanned by the intersecting regions in Figures~\ref{fig7} and~\ref{fig8}
(plus an additional 10~K and 0.05~dex uncertainty in {\teff} and $\log{g}$
for sampling uncertainty);
possible systematic errors of 50--100~K in {\teff} and 0.1--0.25 dex in $\log{g}$
are not included.
The gray circle denotes parameters for 2MASS~0937+2931 assuming [M/H] = $-0.2$ (see $\S$~4.2.2).
\label{fig9}}
\end{figure}

According to brown dwarf evolutionary theory,
the temperature and surface gravity of a brown dwarf
at a given time is directly
related to its mass and age assuming a given composition\footnote{We ignore
for this discussion metallicity effects
in the evolution of a brown dwarf, in addition to other variations
related to rotation, magnetic activity, accretion or binary interaction
that can also modulate the observed parameters and evolution of a brown dwarf.}.
This implies that if {\teff} and $g$ are known, estimates for the latter, more
fundamental physical properties could be derived on an individual basis.
This is demonstrated graphically in Figure~\ref{fig9}, which compares the derived
{\teff} and $\log{g}$ values for 14 of our sources (excluding 2MASS~0939$-$2448
and 2MASS~1114$-$2618) to solar metallicity
evolutionary models from \citet{bur97}.
Table~\ref{tab:mass} lists the range of masses and ages, and the corresponding
radii, derived from this comparison.
Our sample appears to span a broad range of masses
(0.02--0.07 M$_{\sun}$) and ages ($<$1--10 Gyr),
consistent with a random sample drawn from the local Galactic environment.  At
late ages, brown dwarf radii are fairly constant,
so mass and surface gravity are
almost directly related.  Hence, our lowest (highest) surface gravity objects
are also the least (most) massive and youngest (oldest). Note in particular
the placement of 2MASS~0937+2931, which appears to be the most massive
and oldest in the sample.  On the other hand, our analysis suggests that
the low surface gravity T dwarfs SDSS~1758+4633 and 2MASS~2228$-$4310 may
have masses less
than 0.03 M$_{\sun}$ and ages less than 1 Gyr.

How reliable are these estimates?  An independent check of the derived ages can
be made by examining the kinematics of our sample.  Three dimensional
space velocities have not yet been measured for the T dwarfs examined here;
a rough analysis can be made, however, by examining their tangential velocities, $V_{tan}$.
Of the 14 sources in our sample with age estimates, 13 have proper motion measurements
and eight have parallax measurements.  For
2MASS~0034+0523, 2MASS~0050$-$3322, 2MASS~1231+0847 and 2MASS~2228$-$4310
we adopted spectrophotometric distance estimates from \citet{me03e} and \citet{tin05};
a distance for SDSS~1110+0116 was estimated using absolute MKO $M_J$ and $M_K$ magnitude/spectral
type relations from \citet{gol04}.  Dividing the 13 T dwarfs into those with mean
estimated ages less than 2 Gyr (young) and older than 2 Gyr (old), we computed
the mean (${\langle}{V_{tan}}{\rangle}$) and standard deviation (${\sigma}_{V_{tan}}$)
for each age group.
For the young sources, ${\langle}{V_{tan}}{\rangle}$ = 38 km s$^{-1}$
and ${\sigma}_{V_{tan}}$ = 20 km s$^{-1}$, while
for the old sources ${\langle}{V_{tan}}{\rangle}$ = 51 km s$^{-1}$
and ${\sigma}_{V_{tan}}$ = 31 km s$^{-1}$.
The larger mean motion and greater scatter in velocities for the latter group
is consistent with an older mean age.  However, a
rigorous examination of the three dimensional
velocities is required before a conclusive assessment can be made.

\subsection{Independent Mass and Radius Estimates}

The masses and ages of the objects in our sample as derived from the
evolutionary models appear to be reasonable and consistent with
their overall properties.  However,
these values must be considered with
caution as they are susceptible to systematic errors in both the
atmosphere {\em and} interior models.  Disagreements in mass and radius estimates
between these two types of models have been suggested
in a few young systems \citep[however, see Luhman, Stauffer \& Mamajek 2005]{moh04b,clo05},
and such systematic deviations
may be present at late ages as well.
Fortunately, masses and radii can be determined independently of the
evolutionary models for those brown dwarfs with bolometric luminosity measurements,
as follows.

The surface gravity of a solid body, $g = G{\rm M}/{\rm R}^2$,
is applicable for brown dwarf photospheres
since the vertical scaleheight of this region (a few km; Griffith \& Yelle 1999) is insignificant
compared to the radius of the brown dwarf itself
(${\sim}$0.1~R$_{\sun} \equiv 6.95{\times}10^4$ km). Combining this with
the definition of {\teff}, $L = 4{\pi}{\rm R}^2{\sigma}T_{eff}^4$, yields
\begin{eqnarray}
 {\rm M} & = & \frac{Lg}{4{\pi}G{\sigma}T_{eff}^4} \label{eqn:m} \\
   & = & 0.0408\left( \frac{L}{10^{-5} L_{\sun}} \right) \left( \frac{g}{10^5 {\rm cm~s^{-2}}} \right) \left( \frac{T_{eff}}{1000~{\rm K}} \right) ^{-4} {\rm M}_{\sun} \nonumber \\
 \nonumber
 \end{eqnarray}
and
\begin{eqnarray}
 {\rm R} & = & \left( \frac{L}{4{\pi}{\sigma}T_{eff}^4} \right) ^{1/2} \label{eqn:r} \\
   & = & 0.106 \left( \frac{L}{10^{-5} L_{\sun}} \right) ^{1/2} \left( \frac{T_{eff}}{1000~{\rm K}} \right) ^{-2} {\rm R}_{\sun}. \nonumber \\
 \nonumber
 \end{eqnarray}
%where $L_{\sun} = 3.86{\times}10^{33}$ erg s$^{-1}$ and {\msun} = $1.99{\times}10^{33}$ g.
These equations
rely only on the {\teff} and $g$ values obtained from the
spectral models and the measured
luminosities, and not on any
evolutionary model (c.f., Mohanty, Jayawardhana \& Basri 2004)

\begin{figure}
\epsscale{1.0}
\plotone{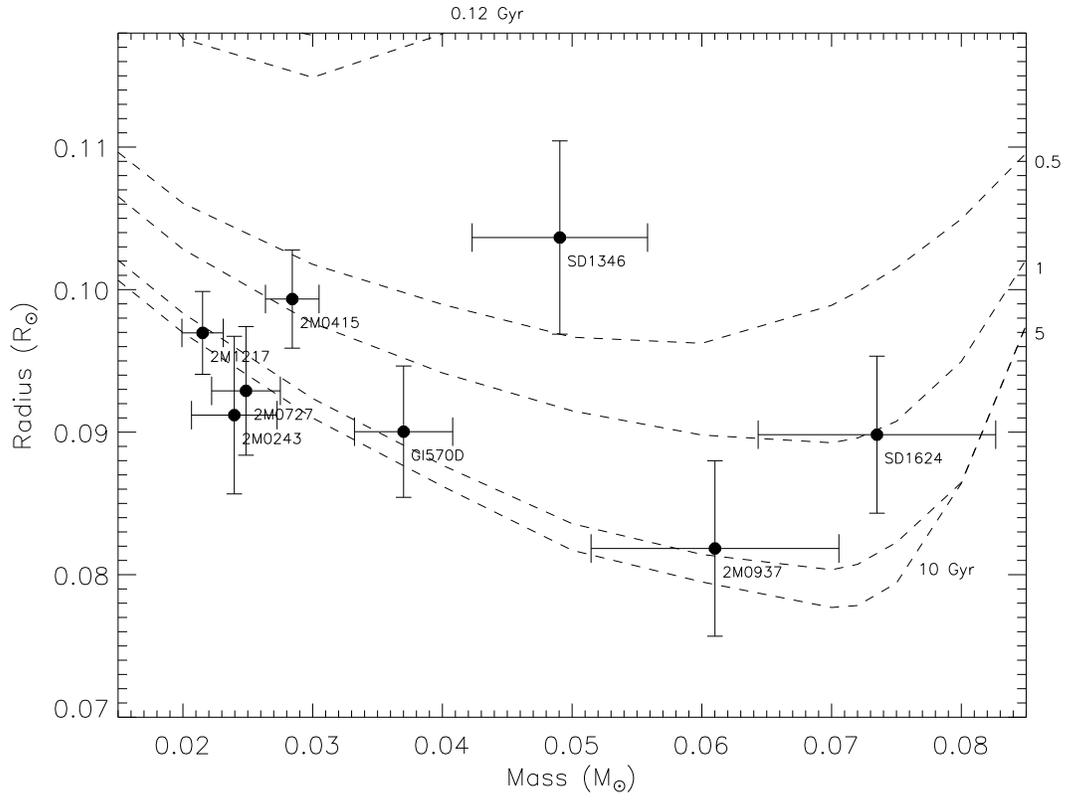}
\caption{Masses and radii derived for T dwarfs with luminosity
determinations from \citet{vrb04}
using Equations~\ref{eqn:m} and~\ref{eqn:r}.  Isochrones (ages in Gyr as labelled) from the
evolutionary models of \citet{bur97} are indicated by dashed lines.
\label{fig10}}
\end{figure}

Luminosities for field brown dwarfs have been compiled by a number of studies
\citep{geb01,dah02,nak04,cus05};
here we focus on the results of
\citet{gol04} and \citet{vrb04}. Seven of the
T dwarfs in our sample have luminosity determinations from these studies;
Golimowski et al.\ also adopt the {\teff} and $\log{g}$ determinations of Gliese~570D
from \citet{geb01}.
The corresponding masses and radii
derived from Eqns.~\ref{eqn:m} and~\ref{eqn:r}, and listed in Table~\ref{tab:mass},
generally agree with those derived from the evolutionary models.
Figure~\ref{fig10}
shows a comparison of mass and radius values derived
from the Vrba et al.\ luminosities
to the \citet{bur97} theoretical isochrones.
With the exception of SDSS~1346$-$0031\footnote{The large radius derived for SDSS~1346$-$0031
could arise if the source is an unresolved binary, although systematic effects cannot
be ruled out.}, derived values lie
between the 1 and 10 Gyr isochrones,
as expected for a Galactic disk sample.
More importantly, features in the theoretical
brown dwarf mass-radius relationship are
reproduced, including the radius minimum
of 0.08~R$_{\sun}$ for larger masses and the trend toward larger radii
for lower mass brown dwarfs \citep{bur97,cha00}.
These agreements are promising, and suggest that brown dwarf
evolutionary tracks are robust at late ages.

Are these values truly independent of the evolutionary models?
Not entirely, since the {\teff} and $\log{g}$ determinations
hinge on the adopted values for Gliese~570D, which themselves are partly
dependent on the evolutionary models.  {\teff} was derived for this source
by \citet{geb01} using
the integrated observed flux (over 0.83--2.52~$\micron$),
a bolometric correction determined from {\em spectral} models
(but consistent with more recent empirical determinations; Golimowski et al.\ 2004),
the measured parallax of the Gliese~570 system, and a radius adopted from evolutionary models.
In the last case, the radii of brown dwarfs at the age of Gliese~570D are predicted to be
roughly constant, varying by less than 20\% for masses of 0.02--0.07~M$_{\sun}$; are
determined by well-understood interior physics; and have been empirically tested
at the low-mass end with transiting extrasolar planets (e.g., \citet{bur04}).
Hence, {\teff} for Gliese~570D can be considered to be empirically robust.
Its surface gravity, on the other hand, was derived solely from evolutionary theory.
However, given the largely constant radii of old brown dwarfs
(R $\approx$ 0.1~R$_{\sun}$), and assuming
a mass in the brown dwarf range (0.02--0.07~M$_{\sun}$) yields $\log{g} = 4.7-5.3$ cm~s$^{-2}$,
consistent with the adopted model-dependent value.  Hence, the adopted {\teff} and $\log{g}$
for Gliese~570D are only weakly tied to evolutionary models, so that
the derived parameters for other T dwarfs can provide, at mimimum, semiempirical tests
of these models.

Two of the objects in our sample, 2MASS~0415$-$0935 and 2MASS~0937+2931,
are worth additional comment as their
luminosities from \citet{gol04} and \citet{vrb04} are significantly discrepant.
As a result, Eqns.~\ref{eqn:m} and~\ref{eqn:r} yield
very different masses and radii for these sources.
Golimowski et al.\ determined luminosities for individual brown dwarfs
by integrating the total measured flux over 0.8--5.0 microns
(for those sources with measured $M$-band photometry),
assuming a Rayleigh-Jeans tail for longer wavelengths (with corrections for molecular absorption
between 4 and 15 $\micron$), and using measured parallaxes.  Vrba et al.\
apply a bolometric correction as a function of spectral type
(computed by Golimowski et al.) to absolute $K$-band magnitudes.  For 2MASS~0415$-$0935,
the slight differences between these methods
yields a lower luminosity from Golimowski et al.\ (by a factor of 3.5),
resulting in a similar mass but a
much smaller radius (0.083$\pm$0.003 R$_{\sun}$) than that derived from the
Vrba et al.\ measure (0.092$\pm$0.004 R$_{\sun}$).
The former estimate is outside of the
\citet{bur97} model parameter space.  Similarly,
Golimowski et al.\ deduce a higher luminosity for 2MASS~0937+2931 than Vrba et al.,
and the corresponding mass (0.118$\pm$0.018 M$_{\sun}$) and radius
(0.114$\pm$0.008 R$_{\sun}$) are well outside of the parameter space encompassed by the
evolutionary models.
In both cases, the Vrba et al.\ luminosities yield values consistent with
the models. This is intriguing, since both 2MASS~0415$-$0935 and 2MASS~0937+2931
have measured $M$-band photometry, and
the corresponding luminosities from Golimowski et al.\ are expected to be more
accurate.  These deviations may indicate systematic errors in the luminosity determinations
of either Golimowski et al.\ or Vrba et al., or in our {\teff} and surface gravity
estimates.  This is not entirely unexpected for the
apparently metal-poor T dwarf 2MASS~0937+2931.  However, in order to assess
the origin of these deviations, and whether
they actually indicate problems in the evolutionary models,
the number and quality of luminosity measurements for low temperature
T dwarfs must clearly be improved.

\section{Discussion}

\subsection{The Temperature Scale of Late-type T Dwarfs}

Disentangling the parameters {\teff} and $g$ for T dwarf spectra enables
a more refined examination of the
{\teff}/spectral type relation for these objects,
a useful function for constraining
atmospheric properties as well as distance estimation.
Typically, these relations are tied to luminosity measurements
and an assumed radius, or range of radii for sources with unknown ages
\citep{dah02,tin03,gol04,vrb04}.
Studies have shown that
late-type T dwarfs with identical spectral types can
exhibit significant differences in their estimated {\teff}s.
We formally recognize this as the
additional influence of surface gravity.

\begin{figure}
\epsscale{0.9}
\plotone{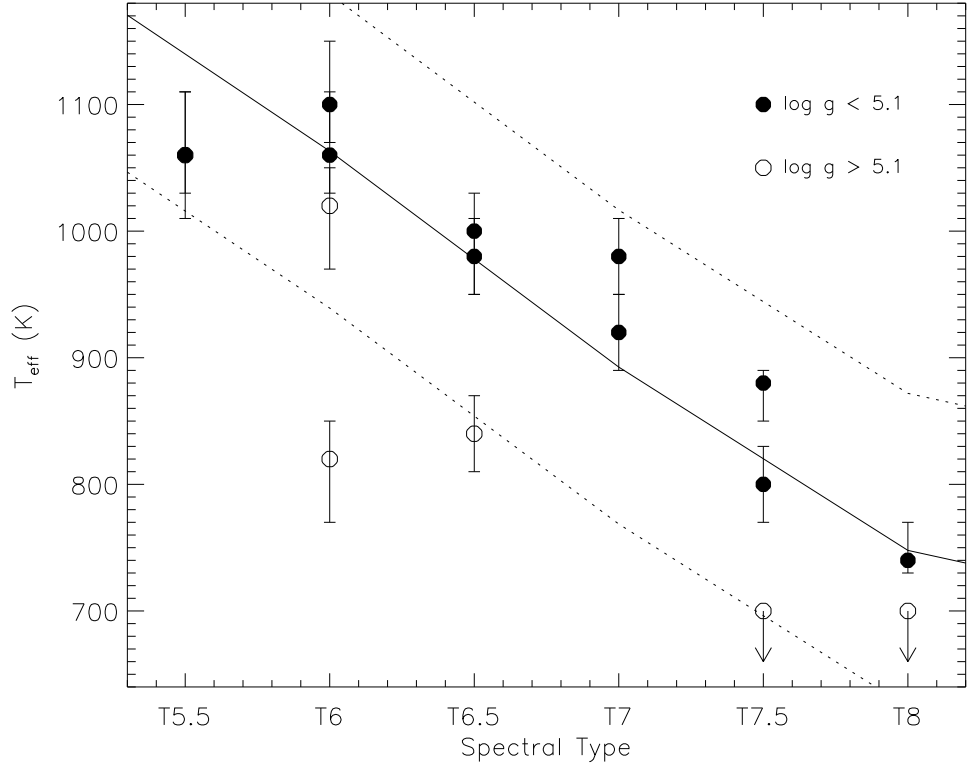}
\caption{Derived {\teff}s for objects in our sample as a function of
near infrared spectral type.  Sources have been segregated into
those with low and moderate surface gravities ($\log{g} \leq 5.1$; filled circles)
and those with high surface gravities ($\log{g} > 5.1$; open circles).
The low and moderate surface gravity objects form a tighter
trend than the full sample, as traced by the \citet{gol04} {\teff}/spectral type relation (solid line;
dashed lines delineate $\pm$124~K uncertainty in the relation);
higher surface gravity objects tend to have lower {\teff}s at a given
spectral type.
\label{fig11}}
\end{figure}

Figure~\ref{fig11}
compares the derived {\teff}s for the sources in our sample to their
spectral types.   Objects with low and moderate surface gravities,
$\log{g} \leq 5.1$, exhibit a tight trend of decreasing {\teff}
with increasing spectral type,
largely consistent with the \citet{gol04} {\teff}/spectral type relation.
Higher surface gravity objects,
in particular 2MASS~0034+0523 and 2MASS~0937+2931, have {\teff}s that
are 150--250~K cooler for their spectral types.  This behavior
can be understood by the interplay between {\teff} and $g$ on
the major {\water} bands, the depths of which determine in part T spectral types.
T dwarfs with higher surface gravities have weaker
{\water} bands, and hence earlier spectral types, at a given {\teff};
consequently, they would appear to have lower {\teff}s for
a given spectral type.  This gravity trend is also present
when comparing luminosities, as the oldest, most massive
brown dwarfs (which have the highest surface gravities)
can have radii that are 10-15\% smaller
than 1--3~Gyr brown dwarfs \citep{bur97}.  Coupled with 10--20\%
lower {\teff}s, old brown dwarfs can be up to three times fainter
than their younger field counterparts.  This is precisely the
deviation \citet{vrb04}
finds in the luminosity of 2MASS~0937+2931 as compared to other T6 dwarfs
(Golimowski et al.\ 2004 find a somewhat smaller deviation).
Hence, luminosity and/or {\teff} measurements for a consistently classified sample
could provide a means of segregating young and old systems.

\subsection{Improving Temperature and Surface Gravity Determinations}

The method outlined here is in some sense a response to the current limitations
of the spectral models.  As the models improve in accuracy,
direct spectral comparisons should eventually be sufficient to determine the physical parameters
of field brown dwarfs.  On the other hand, our spectral index comparison method could also be improved
by using additional empirical calibrators such as Gliese~570D.
Additional calibrators would enable higher order corrections to the model ratios,
reducing systematic effects.
Such empirical constraints can be taken one step
further: a sufficiently sampled parameter space of calibrator
sources could enable the determination of {\teff} and $g$ values
{\em independent of the spectral models}.  To do this,
several more companion brown dwarfs with independent age and metallicity
determinations, and/or binary (particularly eclipsing) systems with measured
orbital parameters, are required.  While Gliese~570D is currently the only such calibrator source
in the late T dwarf regime, three closely-separated late-type T dwarf binaries
have been identified for which
mass measurements are feasible \citep[Burgasser et al.\ in prep.]{me03b};
and one T dwarf binary, Epsilon Indi Bab \citep{sch03,mcc04} is also a companion
to a nearby 0.8--2~Gyr K5 V star.
There are also several ongoing searches for wide brown dwarf companions
to nearby and young stars (e.g., Chauvin et al.\ 2004, 2005; Neuh\"{a}user et al.\ 2005).
Identification of several such calibration stars would
provide an empirical ladder for determining the
physical properties of field brown dwarfs, and a critical
test for both spectral and evolutionary models.

Can we also extend this technique to earlier spectral types; e.g., L dwarfs
and early-type T dwarfs?  In these regimes, spectral energy distributions are strongly
affected by photospheric condensates, giving rise to what many consider to be
a fourth ``dust'' parameter
(e.g., $f_{sed}$ in \citet{ack01} and \citet{mar02}; T$_{cr}$ in \citet{tsu05};
$a_0$ in \citet{bur05})
which may vary with the photospheric gas properties or
other secondary effects, such as rotation.  In principle, the method outlined
here could be extended into the L and early T dwarf regime by incorporating
this fourth parameter, employing suitable dust cloud models and enlarging
the sample of empirical calibration stars.  In practice, this approach may prove
more difficult as brown dwarf cloud formation
remains a poorly understood process (e.g., Helling et al.\ 2004).  Nevertheless, {\teff},
$g$ and dust content determinations would be particularly useful for understanding
the unusual transition where L and T dwarfs when photospheric dust is rapidly depleted
\citep{dah02,me02c,kna04}.

\subsection{Applications of the Method}

The ability to determine masses and ages for individual brown dwarfs is
clearly a boon to statistical studies of these objects, particularly
the Galactic substellar MF and formation history.  Constraints on these
fundamental relations have largely been statistical in nature
because of the difficulty
in determining masses and ages for field sources \citep{rei99,cha03,me04a,all05}.
Using the method described here, both distributions can be built up directly from
individual sources in a sample, assuming that careful consideration is made of
selection effects.  While current samples of late-type T dwarfs are too small for
a robust analysis of this kind \citep{me02a,kna04}, searches for cold brown dwarfs
from wide and deep near infrared surveys such
as the UKIRT Infrared Deep Sky Survey\footnote{See http://www.ukidss.org/index.html.}
\citep{war02} may make MF and age distribution measurements feasible in the near future.

In the more immediate term, {\teff} and $g$ measurements, and corresponding
mass and age estimates, are useful for identifying and characterizing
young, low mass objects in young star forming regions.
According to current evolutionary models,
T dwarfs with ages $\lesssim$~10~Myr can have masses of only a few
Jupiter masses (M$_{Jup}$).
Several young cluster brown dwarf candidates have been identified
in recent years, including the late-type T dwarf
S Orionis 70 \citep{zap02,mar03}.  Direct comparison of spectral data
to theoretical models have suggested a very low surface gravity for this source,
$\log{g} \approx 3.5-4.0$, indicative of a young, very low mass ($\sim$3~M$_{Jup}$)
brown dwarf which has not yet fully contracted.  However, such direct spectral
comparison has been shown to be flawed for late T spectral types in general \citep{me04c}.
A calibrated spectral ratio comparison could provide a more robust assessment of
the physical properties of this and other low mass candidates, and verify
their cluster membership.  Furthermore, with luminosity measurements,
independent determinations of mass and age would provide semiempirical
constraints on the evolutionary models from which mass estimates are currently
derived.  Such independent assessments are necessary to validate the existence of
so-called ``planetary-mass''
brown dwarfs in these young star forming regions.

\section{Summary}

We have devised a method for measuring the effective temperatures
and surface gravities for the lowest luminosity T-type brown dwarfs,
by the comparison of calibrated spectral ratios measured on
low resolution near infrared spectral data and theoretical models.
Using this method, we have derived {\teff} and $g$ estimates for 14 T5.5--T8 field brown dwarfs,
and a subsolar metallicity estimate for the peculiar T dwarf 2MASS~0937+2931.
Two other sources in our sample, 2MASS~0939$-$2448
and 2MASS~1114$-$2618, appear to have {\teff} $\lesssim$ 700~K, and
are potentially the coldest brown dwarfs currently known.
We find no evidence of systematic effects in our method, and the
agreement of our {\teff} determinations with prior studies suggests
that our results are robust.  We also find that the scatter
observed in {\teff}/spectral type relations likely arises from surface
gravity effects, as higher surface gravity objects have a lower
{\teff} at a given spectral type.
Masses, radii and ages are derived for
objects in our sample using the \citet{bur97} evolutionary models, while
independent mass and radius determinations are made for eight T dwarfs
with luminosity measurements.  Broad agreement between these values
suggests that current brown dwarf evolutionary models are accurate at late ages,
although this must be verified through improved luminosity determinations.

The comparative technique described here is a useful tool
for determining the physical properties of the lowest-luminosity brown dwarfs,
making efficient use of low resolution, and therefore more sensitive, spectroscopy.
As such, it is a promising method for characterizing large samples arising from deep surveys,
enabling a direct determination of the Galactic substellar MF and formation history.
While our method remains tied to the current generation of spectral models,
susceptible to persistent opacity deficiencies, the increased use of fiducial
calibrators will ultimately enable a wholly empirical approach, allowing
critical tests of atmospheric and evolutionary theories in addition to the characterization
of individual brown dwarfs in the vicinity of the Sun.

\acknowledgments

We thank our telescope operators, Dave Griep and Paul Sears,
and instrument specialist John Rayner, for their
support during the SpeX observations.  We also acknowledge
extremely helpful comments from our referee, Gibor Basri, and editor,
James Liebert, which allowed us to substantially improve upon our original manuscript.
A.\ Burgasser thanks P.\ Hauschildt for making electronic versions
of their COND models available for analysis.
A.\ Burrows acknowledges support under NASA grant NNG04GL22G.
This publication makes use of data from the Two
Micron All Sky Survey, which is a joint project of the University
of Massachusetts and the Infrared Processing and Analysis Center,
funded by the National Aeronautics and Space Administration and
the National Science Foundation.
2MASS data were obtained from
the NASA/IPAC Infrared Science Archive, which is operated by the
Jet Propulsion Laboratory, California Institute of Technology,
under contract with the National Aeronautics and Space
Administration.
The theoretical material is based upon work enabled by the National Aeronautics and
Space Administration through the NASA Astrobiology Institute under
Cooperative Agreement No.\ CAN-02-OSS-02 issued through the Office of Space Science.
The authors wish to recognize and acknowledge the very significant cultural role and
reverence that the summit of Mauna Kea has always had within the indigenous Hawaiian
community.  We are most fortunate to have the opportunity to conduct observations
from this sacred mountain.

Facilities: \facility{IRTF}{:~SpeX}.

\clearpage

\clearpage

\clearpage

\clearpage

\clearpage

\clearpage

\begin{deluxetable}{llcccccccl}
\tabletypesize{\scriptsize}
%\rotate
\tablecaption{T Dwarf Sample\label{tab:sample}}
\tablewidth{0pt}
\tablehead{
 & & \multicolumn{2}{c}{J2000 Coordinates\tablenotemark{a}} &
 \multicolumn{3}{c}{2MASS Photometry} \\
\cline{3-4} \cline{5-7}
\colhead{Name} &
\colhead{SpT} &
\colhead{$\alpha$} &
\colhead{$\delta$} &
\colhead{$J$} &
\colhead{$H$} &
\colhead{$K_s$} &
\colhead{$\pi$} &
\colhead{$\mu$} &
\colhead{Ref\tablenotemark{b}} \\
 & & & & & & &
\colhead{(mas)} &
\colhead{($\arcsec$ yr$^{-1}$)} \\
\colhead{(1)} &
\colhead{(2)} &
\colhead{(3)} &
\colhead{(4)} &
\colhead{(5)} &
\colhead{(6)} &
\colhead{(7)} &
\colhead{(8)} &
\colhead{(9)} &
\colhead{(10)}  \\
}
\startdata
2MASS J00345157+0523050 & T6.5 & 00$^h$34$^m$51$\fs$57 & +05$\degr$23$\arcmin$05$\farcs$0 &  15.54$\pm$0.05 & 15.44$\pm$0.08 & $>$ 16.2 & \nodata & 0.68$\pm$0.06 & 1 \\
2MASS J00501994$-$3322402 & T7 & 00$^h$50$^m$19$\fs$94 & $-$33$\degr$22$\arcmin$40$\farcs$2 &  15.93$\pm$0.07 & 15.84$\pm$0.19 & 15.24$\pm$0.19 & \nodata & 1.5$\pm$0.1 & 2 \\
2MASS J02431371$-$2453298 & T6 & 02$^h$43$^m$13$\fs$71 & $-$24$\degr$53$\arcmin$29$\farcs$8 & 15.38$\pm$0.05 & 15.14$\pm$0.11 & 15.22$\pm$0.17 & 94$\pm$4 &  0.355$\pm$0.004 & 3,4 \\
2MASS J04151954$-$0935066 & T8 & 04$^h$15$^m$19$\fs$54 & $-$09$\degr$35$\arcmin$06$\farcs$6 &  15.70$\pm$0.06 & 15.54$\pm$0.11 & 15.43$\pm$0.20 & 174$\pm$3 & 2.255$\pm$0.003 &  3,4 \\
2MASS J07271824+1710012 & T7 & 07$^h$27$^m$18$\fs$24 & +17$\degr$10$\arcmin$01$\farcs$2 &  15.60$\pm$0.06 & 15.76$\pm$0.17 & 15.56$\pm$0.19 & 110$\pm$2 & 1.297$\pm$0.005 & 3,4 \\
2MASS~J09373487+2931409 & T6p & 09$^h$37$^m$34$\fs$87 & +29$\degr$31$\arcmin$40$\farcs$9 & 14.65$\pm$0.04 & 14.70$\pm$0.07 & 15.27$\pm$0.13 & 163$\pm$4 & 1.622$\pm$0.007 &  3,4 \\
2MASS~J09393548$-$2448279 & T8 & 09$^h$39$^m$35$\fs$48 & $-$24$\degr$48$\arcmin$27$\farcs$9 & 15.98$\pm$0.11 & 15.80$\pm$0.15 & $>$ 16.6 & \nodata & 1.15$\pm$0.06 & 2 \\
SDSS J111010.01+011613.1 & T5.5 & 11$^h$10$^m$10$\fs$01 & +01$\degr$16$\arcmin$13$\farcs$0 &  16.34$\pm$0.12 & 15.92$\pm$0.14 & $>$ 15.1 & \nodata & 0.34$\pm$0.10 & 2,5 \\
2MASS~J11145133$-$2618235 & T7.5 & 11$^h$14$^m$51$\fs$33 & $-$26$\degr$18$\arcmin$23$\farcs$5 & 15.86$\pm$0.08 & 15.73$\pm$0.12 & $>$ 16.1 & \nodata & 3.05$\pm$0.04 &  2 \\
2MASS~J1217110$-$0311131 & T7.5 & 12$^h$17$^m$11$\fs$10 & $-$03$\degr$11$\arcmin$13$\farcs$1 & 15.86$\pm$0.06 & 15.75$\pm$0.12 & $>$ 15.9 & 91$\pm$2 & 1.0571$\pm$0.0017 &  6,7 \\
2MASS J12314753+0847331 & T5.5 & 12$^h$31$^m$47$\fs$53 & +08$\degr$47$\arcmin$33$\farcs$1 &  15.57$\pm$0.07 & 15.31$\pm$0.11 & 15.22$\pm$0.20 & \nodata & 1.61$\pm$0.07 & 1,2,8 \\
SDSS~J134646.45$-$003150.4 & T6.5 & 13$^h$46$^m$46$\fs$34 & $-$00$\degr$31$\arcmin$50$\farcs$1 & 16.00$\pm$0.10 & 15.46$\pm$0.12 & 15.77$\pm$0.27 & 68$\pm$2 & 0.516$\pm$0.003 & 9,7 \\
Gliese~570D & T7.5 & 14$^h$57$^m$14$\fs$96 & $-$21$\degr$21$\arcmin$47$\farcs$7 & 15.32$\pm$0.05 & 15.27$\pm$0.09 & 15.24$\pm$0.16 & 169.3$\pm$1.7 & 2.012$\pm$0.004 &  10,11 \\
SDSS~J162414.37+002915.6 & T6 & 16$^h$24$^m$14$\fs$36 & +00$\degr$29$\arcmin$15$\farcs$8 & 15.49$\pm$0.05 & 15.52$\pm$0.10 & $>$ 15.5 & 92$\pm$2 &  0.3832$\pm$0.0019 & 12,13 \\
SDSS J175805.46+463311.9 & T6.5 & 17$^h$58$^m$05$\fs$45 & +46$\degr$33$\arcmin$09$\farcs$9 &  16.15$\pm$0.09 & 16.25$\pm$0.22 & 15.47$\pm$0.19 & \nodata & \nodata & 8 \\
2MASS~J22282889$-$4310262 & T6 & 22$^h$28$^m$28$\fs$89 & $-$43$\degr$10$\arcmin$26$\farcs$2 & 15.66$\pm$0.07 & 15.36$\pm$0.12 & 15.30$\pm$0.21 & \nodata & 0.31$\pm$0.03 &  14 \\
\enddata
\tablenotetext{a}{Right Ascension ($\alpha$) and declination ($\delta$)
at equinox J2000 from the 2MASS All Sky Data Release Point Source Catalog \citep{cut03}.}
\tablenotetext{b}{Discovery, parallax and proper motion references.}
\tablerefs{(1) \citet{me04b}; (2) \citet{tin05}; (3) \citet{me02a}; (4) \citet{vrb04};
(5) \citet{geb02}; (6) \citet{me99}; (7) \citet{tin03}; (8) \citet{kna04};
(9) \citet{tsv00}; (10) \citet{me00a}; (11) HIPPARCOS \citep{esa97};
(12) \citet{str99}; (13) \citet{dah02}; (14) \citet{me03e}.}
\end{deluxetable}

\begin{deluxetable}{lcccll}
\tabletypesize{\footnotesize}
\tablecaption{Log of New SpeX Observations.\label{tab:spexlog}}
\tablewidth{0pt}
\tablehead{
\colhead{Source} &
\colhead{UT Date} &
\colhead{$\sec{z}$} &
\colhead{$t_{int}$ (s)} &
\colhead{Calibrator Star} &
\colhead{SpT} \\
\colhead{(1)} &
\colhead{(2)} &
\colhead{(3)} &
\colhead{(4)} &
\colhead{(5)} &
\colhead{(6)}  \\
}
\startdata
2MASS~0050$-$3322 & 2004 Sep 07 & 1.69 & 1440 & HD 225200 & A0 V \\
2MASS~0727+1710  & 2004 Mar 10 & 1.00 & 1280 & HD 56386 & A0 Vn \\
2MASS~0937+2931  & 2004 Mar 11 & 1.02 & 720 & HD 89239 & A0 V \\
2MASS~0939$-$2448 &  2004 Mar 12 & 1.43 & 1080 & HD 81694  & A0 V \\
SDSS~1110+0116 &  2004 Mar 11 & 1.14 & 1800 & HD 97585 & A0 V   \\
2MASS~1114$-$2618 &  2004 Mar 12 & 1.75 & 1080 & HD 98949  & A0 V \\
2MASS~1217$-$0311 &  2004 Mar 11 & 1.13 & 720 & HD 109309 & A0 V  \\
SDSS~1346$-$0031 &  2004 Mar 12 & 1.09 & 720 & HD 116960 & A0 V  \\
SDSS~1624+0029 &  2004 Mar 12 & 1.06 & 720 & HD 136831 & A0 V \\
SDSS~1758+4633 &  2004 Jul 23 & 1.22 & 720 & HD 158261 & A0 V  \\
\enddata
\end{deluxetable}

\begin{deluxetable}{llccccc}
\tabletypesize{\footnotesize}
\tablecaption{Spectral Ratios.\label{tab:ratios}}
\tablewidth{0pt}
\tablehead{
\colhead{Source} &
\colhead{SpT} &
\colhead{{\water}$-J$} &
\colhead{{\water}$-H$} &
\colhead{$K/J$} &
\colhead{$K/H$} &
\colhead{$Y/J$} \\
\colhead{(1)} &
\colhead{(2)} &
\colhead{(3)} &
\colhead{(4)} &
\colhead{(5)} &
\colhead{(6)} &
\colhead{(7)}  \\
}
\startdata
2MASS 0034+0523 & T6.5 & 0.103 & 0.229 & 0.100 & 0.218 & 0.456 \\
2MASS 0050$-$3322 & T7 & 0.104 & 0.266 & 0.180 & 0.388 & 0.363 \\
2MASS 0243$-$2453 & T6 & 0.145 & 0.297 & 0.197 & 0.406 & 0.444 \\
2MASS 0415$-$0935 & T8 & 0.041 & 0.183 & 0.131 & 0.255 & 0.382 \\
2MASS 0727+1710 & T7 & 0.085 & 0.224 & 0.164 & 0.351 & 0.427 \\
2MASS 0937+2931 & T6p & 0.151 & 0.316 & 0.076 & 0.174 & 0.539 \\
2MASS 0939$-$2448 & T8 & 0.038 & 0.149 & 0.059 & 0.117 & 0.493 \\
SDSS 1110+0116 & T5.5 & 0.152 & 0.303 & 0.217 & 0.379 & 0.497 \\
2MASS 1114$-$2618 & T7.5 & 0.039 & 0.177 & 0.076 & 0.150 & 0.482 \\
2MASS 1217$-$0311 & T7.5 & 0.066 & 0.207 & 0.179 & 0.366 & 0.374 \\
2MASS 1231+0847 & T5.5 & 0.181 & 0.271 & 0.157 & 0.328 & 0.451 \\
SDSS 1346$-$0031 & T6.5 & 0.131 & 0.278 & 0.156 & 0.351 & 0.430 \\
Gliese 570D & T7.5 & 0.063 & 0.198 & 0.116 & 0.253 & 0.397 \\
SDSS 1624+0029 & T6 & 0.154 & 0.280 & 0.142 & 0.311 & 0.422 \\
SDSS 1758+4633 & T6.5 & 0.101 & 0.247 & 0.200 & 0.411 & 0.400 \\
2MASS 2228$-$4310 & T6 & 0.157 & 0.293 & 0.204 & 0.440 & 0.383 \\
\cline{1-7}
Corrections\tablenotemark{a} & \nodata & 1.173 & 1.567 & 0.952 & 1.064 & 0.883 \\
\enddata
\tablenotetext{a}{Scale factors applied to model ratios to bring them into agreement
with measurements for Gliese~570D assuming {\teff} = 800~K and $\log{g}$ = 5.1;
See $\S$~4.1.}
\end{deluxetable}

\begin{deluxetable}{llcccccc}
\tabletypesize{\footnotesize}
\tablecaption{Derived Physical Parameters For T Dwarfs.\label{tab:tg}}
\tablewidth{0pt}
\tablehead{
 & & & & & \multicolumn{3}{c}{Published Values} \\
\cline{6-8}
\colhead{Source} &
\colhead{SpT} &
\colhead{\teff} &
\colhead{$\log{g}$} &
\colhead{[M/H]} &
\colhead{\teff\tablenotemark{a}} &
\colhead{\teff\tablenotemark{b}} &
\colhead{$\log{g}$\tablenotemark{c}} \\
\colhead{} &
\colhead{} &
\colhead{(K)} &
\colhead{(cm s$^{-2}$)} &
\colhead{} &
\colhead{(K)} &
\colhead{(K)} &
\colhead{(cm s$^{-2}$)} \\
\colhead{(1)} &
\colhead{(2)} &
\colhead{(3)} &
\colhead{(4)} &
\colhead{(5)} &
\colhead{(6)} &
\colhead{(7)} &
\colhead{(8)}  \\
}
\startdata
2MASS~0034+0523 & T6.5 & 820--860 & 5.4--5.5 & $-0.2{\sim}0$ & \nodata  & \nodata  & \nodata \\
2MASS~0050$-$3322 & T7 & 960--1000 & 4.8--5.0 & 0  & \nodata & \nodata  & \nodata \\
2MASS~0243$-$2453 & T6 & 1040--1100 & 4.8--5.1 & 0 & 825--1150 & 950--1170 & 4.5 \\
2MASS~0415$-$0935 & T8 & 740--760 & 4.9--5.0 & 0 & 600--750 & 690--850 & 5.0 \\
2MASS~0727+1710 & T7 & 900--940 & 4.8--5.0 & 0 & 725--950 & 830--1020 & 4.5  \\
2MASS~0937+2931 & T6p & 780--840\tablenotemark{d} & 5.3--5.5\tablenotemark{d} & $-0.4{\sim}-0.1$ & 725--1000 & 700--850 & 5.5  \\
2MASS~0939$-$2448 & T8 & $\lesssim$700 & \nodata & \nodata & \nodata & \nodata & \nodata \\
SDSS~1110+0116 & T5.5 & 1020--1100 & 4.9--5.2 & 0 & \nodata & \nodata & 4.5  \\
2MASS~1114$-$2618 & T7.5 & $\lesssim$700 & \nodata & \nodata & \nodata & \nodata & \nodata  \\
2MASS~1217$-$0311 & T7.5 & 860--880 & 4.7--4.9 & 0 & 725--925 & 820--1000 & 4.5  \\
2MASS~1231+0847 & T5.5 & 1040--1100 & 5.2--5.5 & 0 & \nodata & \nodata & 5.0 \\
SDSS~1346$-$0031 & T6.5 & 960--1020 & 5.0--5.2 & 0 & 875--1200 & 950--1180 & 4.5  \\
Gliese~570D & T7.5 & 780--820 & 5.1 & 0 & 784--824\tablenotemark{e} & \nodata & 5.0--5.3\tablenotemark{e}  \\
SDSS~1624+0029 & T6 & 980--1040 & 5.3--5.4 & 0 & 800--1100 & 920--1100 & 5.0 \\
SDSS~1758+4633 & T6.5 & 960--1000 & 4.7--4.9 & 0 &  \nodata & \nodata & 4.5 \\
2MASS~2228$-$4310 & T6 & 1080--1140 & 4.6--5.0 & 0 & \nodata & \nodata & \nodata \\
\enddata
\tablenotetext{a}{{\teff} from \citet{gol04}.}
\tablenotetext{b}{{\teff} from \citet{vrb04}.}
\tablenotetext{c}{$\log{g}$ from \citet{kna04}.}
\tablenotetext{d}{{\teff} and $\log{g}$ for [M/H] = $-0.2$. See $\S$~4.2.2.}
\tablenotetext{e}{{\teff} and $\log{g}$ from \citet{geb01}.}
\end{deluxetable}

\begin{deluxetable}{llccccccccc}
\tabletypesize{\scriptsize}
%\rotate
\tablecaption{Estimated Masses, Radii and Ages for T Dwarfs.\label{tab:mass}}
\tablewidth{0pt}
\tablehead{
 & & \multicolumn{3}{c}{Evolutionary Models\tablenotemark{a}} &
\multicolumn{3}{c}{Golimowski et al. Luminosities} &
\multicolumn{3}{c}{Vrba et al. Luminosities} \\
\cline{3-5} \cline{6-11}
\colhead{Source} &
\colhead{SpT} &
\colhead{M} &
\colhead{R} &
\colhead{Age} &
\colhead{$\log{L_{bol}}$} &
\colhead{M} &
\colhead{R} &
\colhead{$\log{L_{bol}}$} &
\colhead{M} &
\colhead{R} \\
\colhead{} &
\colhead{} &
\colhead{(M$_{\sun}$)} &
\colhead{(R$_{\sun}$)} &
\colhead{(Gyr)} &
\colhead{(L$_{\sun}$)} &
\colhead{(M$_{\sun}$)} &
\colhead{(R$_{\sun}$)} &
\colhead{(L$_{\sun}$)} &
\colhead{(M$_{\sun}$)} &
\colhead{(R$_{\sun}$)} \\
\colhead{(1)} &
\colhead{(2)} &
\colhead{(3)} &
\colhead{(4)} &
\colhead{(5)} &
\colhead{(6)} &
\colhead{(7)} &
\colhead{(8)} &
\colhead{(9)} &
\colhead{(10)} &
\colhead{(11)}  \\
}
\startdata
2MASS~0034+0523 & T6.5 & 0.039--0.055 & 0.081--0.090 & 3.4--6.9 & \nodata & \nodata & \nodata & \nodata & \nodata & \nodata \\
2MASS~0050$-$3322 & T7 & 0.022--0.043 & 0.090--0.104 & 0.5--2.5 & \nodata & \nodata & \nodata & \nodata & \nodata & \nodata \\
2MASS~0243$-$2453 & T6 & 0.024--0.041 & 0.092--0.106 & 0.4--1.7 & $-5.08{\pm}0.06$ & 0.021$\pm$0.002 & 0.086$\pm$0.005 & $-5.03{\pm}0.10$ & 0.023$\pm$0.003 & 0.091$\pm$0.005 \\
2MASS~0415$-$0935 & T8 & 0.022--0.044 & 0.085--0.101 & 1.0--4.9 & $-5.73{\pm}0.05$ & 0.020$\pm$0.001 & 0.083$\pm$0.002 & $-5.18{\pm}0.10$ & 0.028$\pm$0.002 & 0.099$\pm$0.003 \\
2MASS~0727+1710 & T7 & 0.022--0.035 & 0.093--0.104 & 0.5--2.2 & $-5.35{\pm}0.05$ & 0.020$\pm$0.002 & 0.083$\pm$0.003 & $-5.26{\pm}0.10$ & 0.024$\pm$0.002 & 0.092$\pm$0.004 \\
2MASS~0937+2931 & T6p & 0.047--0.063 & 0.078--0.084 & 5.5--10 & $-5.28{\pm}0.05$ & 0.118$\pm$0.018 & 0.114$\pm$0.008 & $-5.57{\pm}0.08$ & 0.061$\pm$0.009 & 0.081$\pm$0.006 \\
%2MASS~0939$-$2448 & T8 & 0.034--0.061 & 0.079--0.092 & 3.7--10 & \nodata & \nodata & \nodata & \nodata & \nodata & \nodata \\
SDSS~1110+0116 & T5.5 & 0.028--0.050 & 0.087--0.101 & 0.5--3.4 & \nodata & \nodata & \nodata & \nodata & \nodata & \nodata \\
%2MASS~1114$-$2618 & T7.5 & 0.029--0.053 & 0.081--0.099 & 0.8--10 & \nodata & \nodata & \nodata & \nodata & \nodata & \nodata \\
2MASS~1217$-$0311 & T7.5 & 0.022--0.038 & 0.091--0.103 & 0.7--2.9 & $-5.32{\pm}0.05$ & 0.020$\pm$0.001 & 0.094$\pm$0.002 & $-5.30{\pm}0.09$ & 0.021$\pm$0.001 & 0.096$\pm$0.002 \\
2MASS~1231+0847 & T5.5 & 0.038--0.071 & 0.078--0.093 & 1.6--9.1 & \nodata & \nodata & \nodata & \nodata & \nodata & \nodata \\
SDSS~1346$-$0031 & T6.5 & 0.031--0.056 & 0.082--0.097 & 1.0--4.9 & $-5.00{\pm}0.06$ & 0.051$\pm$0.006 & 0.106$\pm$0.006 & $-5.18{\pm}0.12$ & 0.049$\pm$0.006 & 0.103$\pm$0.006 \\
Gliese~570D & T7.5 & 0.041--0.043 & 0.087--0.089 & 3.7--4.5 & $-5.53{\pm}0.05$\tablenotemark{b} & 0.037$\pm$0.003 & 0.090$\pm$0.004 & \nodata & \nodata & \nodata \\
SDSS~1624+0029 & T6 & 0.054--0.060 & 0.080--0.084 & 4.3--5.8 & $-5.16{\pm}0.05$ & 0.065$\pm$0.007 & 0.084$\pm$0.005 & $-5.11{\pm}0.08$ & 0.073$\pm$0.009 & 0.089$\pm$0.005 \\
SDSS~1758+4633 & T6.5 & 0.019--0.030 & 0.097--0.111 & 0.3--0.9 & \nodata & \nodata & \nodata & \nodata & \nodata & \nodata \\
2MASS~2228$-$4310 & T6 & 0.018--0.034 & 0.096--0.115 & 0.2--0.9 & \nodata & \nodata & \nodata & \nodata & \nodata & \nodata \\
\enddata
\tablenotetext{a}{Estimates derived from the solar metallicity
evolutionary models of \citet{bur97}
and {\teff} and $\log{g}$ determinations from Table~\ref{tab:tg}.}
\tablenotetext{b}{{L}$_{bol}$ from \citet{geb01}.}
\end{deluxetable}

\end{document}